\documentclass[journal]{IEEEtran}

\usepackage{graphicx}
\usepackage{latexsym}
\usepackage{amsfonts}
\usepackage{amssymb}
\usepackage{amsbsy}
\usepackage{amsmath}
\usepackage{multicol}
\usepackage{float}
\usepackage{subfigure}
\usepackage{algorithm}
\usepackage[dvips]{epsfig}

\usepackage{fancyhdr}
\usepackage{epsfig}
\usepackage{threeparttable}
\usepackage{epsf,epsfig}
\usepackage{amsthm}
\usepackage[noadjust]{cite}
\usepackage{dsfont}
\usepackage{color}
\usepackage{enumerate}
\usepackage{bbm}

\usepackage{array}
\makeatletter
\newcommand{\thickhline}{%
    \noalign {\ifnum 0=`}\fi \hrule height 1pt
    \futurelet \reserved@a \@xhline
}
\newcolumntype{"}{@{\hskip\tabcolsep\vrule width 1pt\hskip\tabcolsep}}
\makeatother

\DeclareMathOperator\erf{erf}
\usepackage{amsmath}
\usepackage{algorithm}
\usepackage[noend]{algpseudocode}
\algrenewcommand{\algorithmiccomment}[1]{\hskip3em$\triangleright$ #1}

\makeatletter
\def\BState{\State\hskip-\ALG@thistlm}
\makeatother

\newtheorem{remark}{Remark}

\newtheorem{lemma}{Lemma}
\newtheorem{proposition}{Proposition}
\newtheorem{corollary}{Corollary}

\def\SIR{\mathsf{SIR}}
\def\L{d_l}
\def\W{d_w}

\def\({\left(}
\def\){\right)}
\def\[{\left[}
\def\]{\right]}

\def\nn{\nonumber}

\usepackage{dblfloatfix} 

\def\papertitle{Millimeter-Wave Interference Avoidance via Building-Aware Associations}

\begin{document}

\title{\fontsize{23}{23}\selectfont \papertitle}

\author{Jeemin~Kim,
Jihong~Park\IEEEauthorrefmark{2},
Seunghwan~Kim,
Seong-Lyun~Kim,
Ki~Won~Sung\IEEEauthorrefmark{3},
and~Kwang~Soon~Kim
\thanks{J. Kim, S. Kim, S.-L. Kim, and K. S. Kim are with the School of Electrical and Electronic Engineering, Yonsei University, Seoul, Korea (email: \{jmkim,shkim\}{@ramo.yonsei.ac.kr}, \{slkim,ks.kim\}@yonsei.ac.kr).}
\thanks{\IEEEauthorrefmark{2}J. Park is with the Department of Electronic Systems, Aalborg University, Denmark (email: jihong@es.aau.dk). }
\thanks{\IEEEauthorrefmark{3}K. W. Sung is with Wireless@KTH, KTH Royal Institute of Technology, Kista, Sweden (e-mail: sungkw@kth.se).}}

\maketitle

\begin{abstract}
Signal occlusion by building blockages is a double-edged sword for the performance of millimeter-wave (mmW) communication networks. Buildings may dominantly attenuate the useful signals, especially when mmW base stations (BSs) are sparsely deployed compared to the building density. In the opposite BS deployment, buildings can block the undesired interference. To enjoy only the benefit, we propose a \emph{building-aware association} scheme that adjusts the directional BS association bias of the user equipments (UEs), based on a given building density and the concentration of UE locations around the buildings. The association of each BS can thereby be biased: (i) toward the UEs located against buildings for avoiding interference to other UEs; or (ii) toward the UEs providing their maximum reference signal received powers (RSRPs). The proposed association scheme is optimized to maximize the downlink average data rate derived by stochastic geometry. Its effectiveness is validated by simulation using real building statistics.
\end{abstract}


\IEEEpeerreviewmaketitle

\section{Introduction}

The use of millimeter-wave (mmW) spectrum is a promising way to achieve the 1,000-fold capacity improvement in 5G cellular networks \cite{mmw_intro1,mmw_intro2}.
It is envisaged to provide 20-100 times larger bandwidth, and thereby to resolve the spectrum crunch of the traditional cellular systems using the spectrum below $6$ GHz carrier frequency. Utilizing such abundant mmW spectrum resource is however not free, but needs to pay a price for the significant distance attenuation induced by its vulnerability to signal blockages such as buildings and human bodies \cite{Ericsson15}. 

To compensate the severe distance loss of mmW signals, it is suitable to make the beam mainlobes sharpened and aligned toward the target directions \cite{Ericsson15,beam_coord_mag,JH15}. Due to this strong signal directionality, mmW signals may rarely interfere with each other. Nevertheless, if interference occurs, its impact is detrimental \cite{beam_coord_book}. The mmW beam directions should therefore be carefully decided in order to avoid significant interference.

Motivated by this, in this paper we seek a design for deciding the interference-avoiding mmW beam directions in a downlink urban outdoor scenario. A straightforward solution can be made by using full interference channel information at each base station (BS). This, unfortunately, requires a massive number of channel information exchanges among all BSs in a recursive manner \cite{yang13}. Instead, we focus on the building density within the network region, compared to BS density. The buildings are obviously a major source of mmW signal blockages, and therefore the building density determines how much and how frequently mmW interference occurs. When building density is relatively high, both interfering probability and interference amount become negligibly small. Otherwise, it necessitates a sophisticated beam direction control to avoid interference.

In pursuit of avoiding mmW interference, we suggest a beam control idea making mmW beams steered toward buildings so as to maximize the downlink average data rate. The user equipments (UEs) in front of a building can thereby receive the desired signals, while less incurring interference to the UEs behind the building. Since the locations of UEs and buildings are random, the beam direction control is a challenging task. In fact, too much beam steering toward buildings may significantly increase the leakage of beams that interferes with the UEs adjacent to the buildings. 
Furthermore, the extreme beam steering may lead to no UEs associated within the beam directions and/or to increase the association distances, while unbalancing the association loads among BSs. 
These issues become more critical in an urban outdoor scenario where UEs are concentrated around buildings\footnotemark. 

\footnotetext{People are likely to be gathered around the buildings around which sidewalks, bus stops, and cafe terraces are placed \cite{gehl11,Klaus03,cityreader}.}

\begin{figure*}     
\centering
   \subfigure[Without building-aware association]{
     \includegraphics[width=8cm]{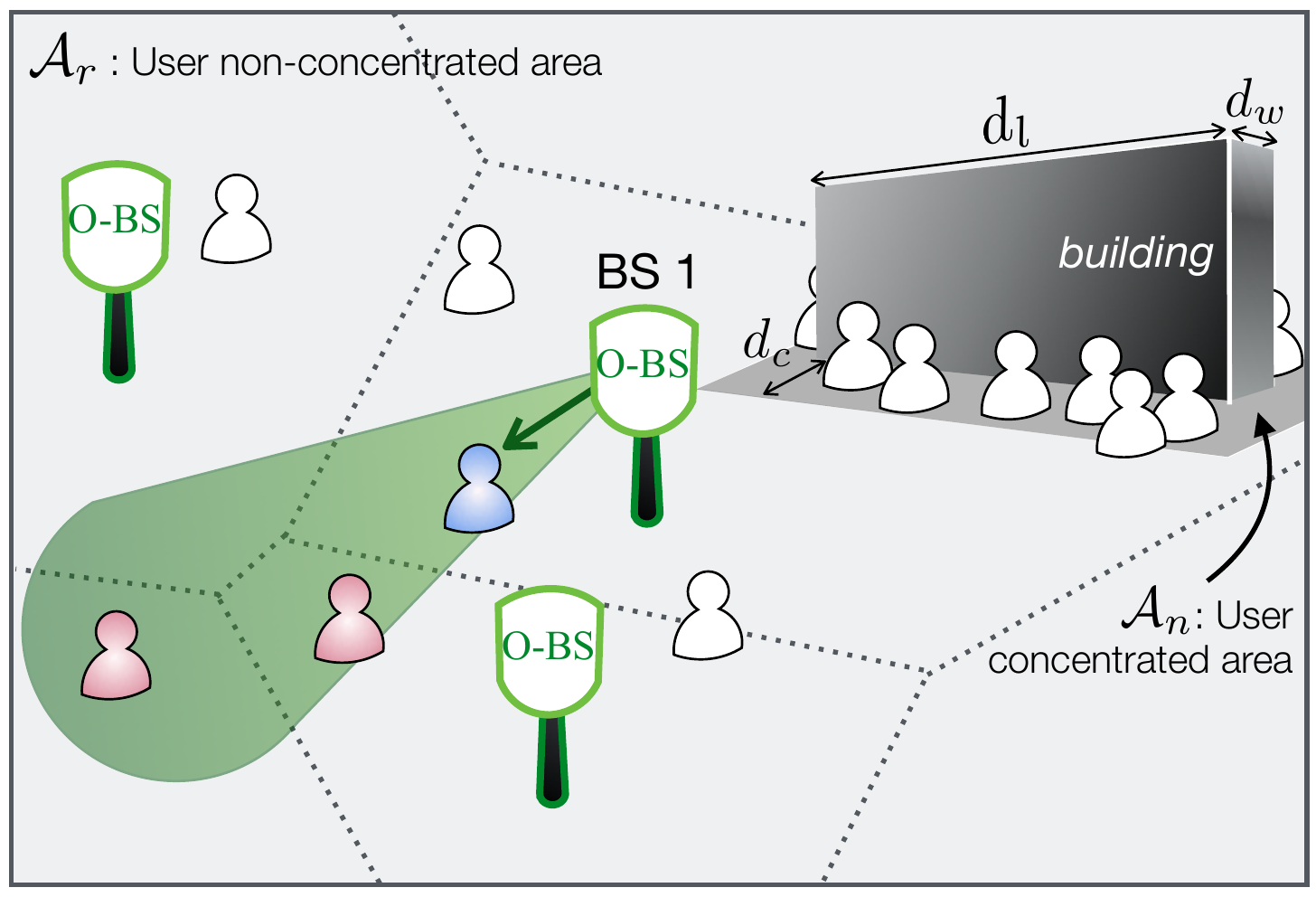}}
   \subfigure[With building-aware association]{\centering
     \includegraphics[width=8cm]{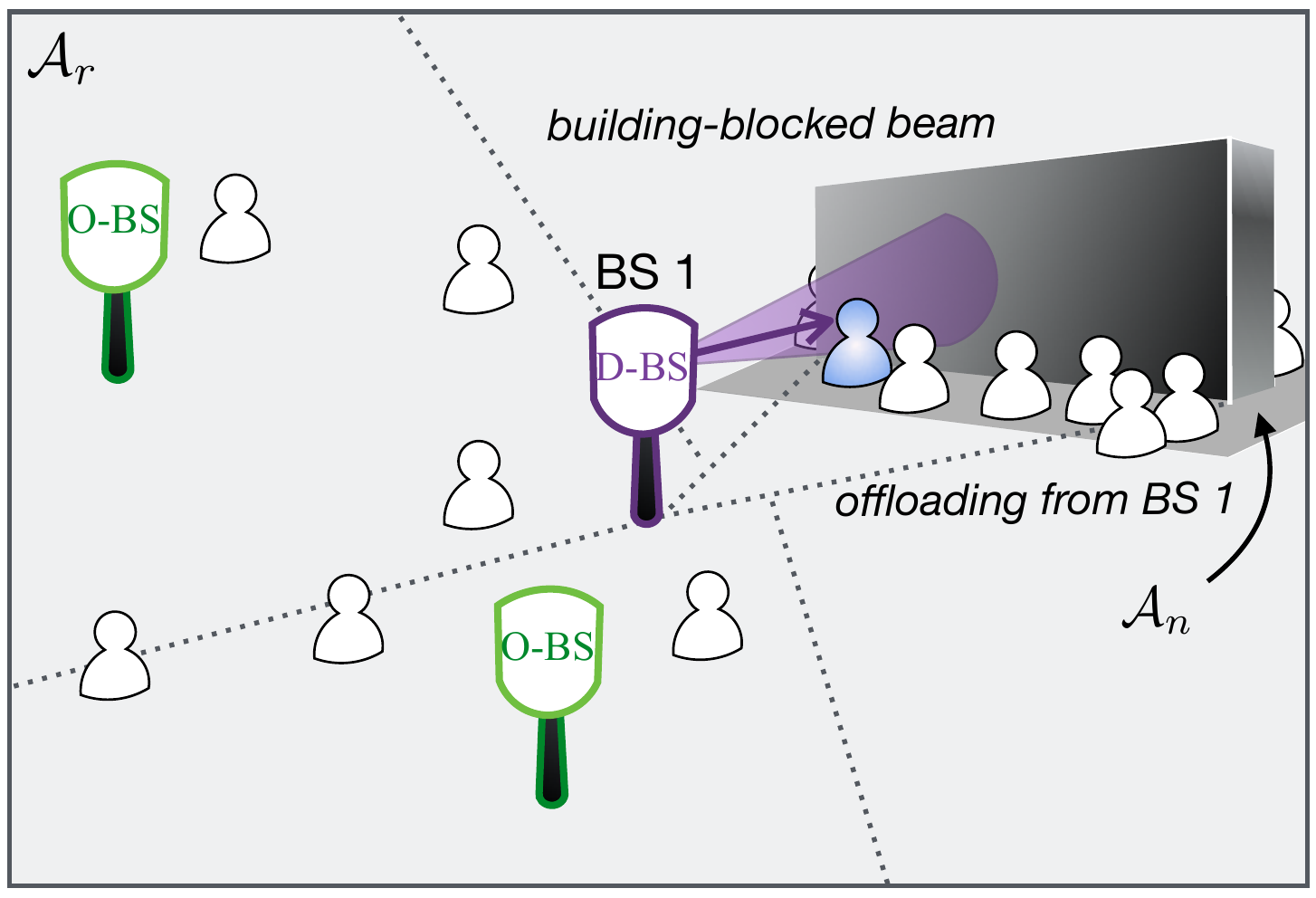} \label{system_b}}
\caption{Illustration of building-aware association algorithm in mmW cellular networks.
UEs are concentrated around the buildings (compare $\mathcal{A}_n$ with $\mathcal{A}_r$) whose average length and width are $d_l$ and $d_w$, respectively.
Without the building-aware association, a signal from BS $1$ interferes with two UEs (see red UEs in (a)).
Such inter-cell interference is removed by the aid of the proposed association which makes BS $1$ transmit signals only toward UEs on the building side (D-BS); \textbf{Interference Decrement}.
Furthermore, by shrinking the association areas of BS $1$, traffic loads are balanced, i.e. the traffic load of D-BSs are offloaded to the BS transmitting signals in every directions (O-BS); \textbf{Load Balancing} (compare (a) with (b)).
} \label{system} 
\end{figure*}

In this paper, we tackle these issues by proposing a \emph{building-aware association} algorithm with a directional association bias $\beta\in[0,1]$ that adjusts the beam steering amount toward buildings (see Fig.~\ref{system}). 
The proposed design is operated based on the full knowledge of each BS's neighboring building locations and sizes, obtained via the map of the network area. 

Specifically, if $\beta=0$, i.e. a baseline without building-aware associations, each BS omni-directionally broadcasts a reference signal as in conventional LTE systems~\cite{rsrpref}. A BS allows the UE associations, ensuring the UEs' maximum average reference signal received powers (RSRPs). Such a BS is hereafter denoted as O-BS.

When $0<\beta\leq1$, with the proposed algorithm, some BSs only allow the associations of the UEs in front of their nearest buildings, hereafter denoted as D-BSs. A BS becomes D-BS if its mainlobe beam is not leaked to the sides of the nearest building, when the beam points at the center of the building. 
The rest of the BSs are set as O-BSs which keep allowing omni-directional UE associations (see Fig.~\ref{Fig:bldecision_a}).

As $\beta$ increases, the ratio of D-BS to O-BS increases. Here, if D-BSs broadcast the reference signals only in the fixed directions toward buildings, some UEs may not receive any reference signals, stuck in the network coverage holes. To prevent this in the proposed algorithm, instead of a BS, each UE omni-directionally broadcasts a reference signal if it does not receive any reference signal, and associates with the BS reporting the maximum RSRP out of all BSs. By so doing, every UE ensures its BS association.

Finally, $\beta$ is optimized so as to maximize the average downloading rate of a randomly selected UE, i.e. a typical UE, while coping with the increases in association distances and load unbalancing due to the directional association biasing.

\subsection{Related Works}

The inter-BS beam coordination has a great potential to improve network capacity \cite{beam_coord_mag}, \cite{bae14,marzi15,boccardi16}.
In \cite{bae14} and \cite{marzi15}, the coordination algorithm where three adjacent BSs coordinate their beams has been investigated.
However, in a real network, the amount of channel information sharing between cooperative BSs is limited due to the backhaul capacity.
To reduce the backhaul burden, a centralized beam coordination has also been studied where the actions are decided by a central entity \cite{boccardi16}.
Although such a centralized scheme reduces the channel information exchange, its viability still relies on the latency of the exchanged information.
Furthermore, in some network scenarios where the central entity is absent, e.g. spectrum sharing among different radio access technologies, such beam coordination cannot be utilized.

Traditionally, the blockage-vulnerable nature of mmW signals in most prior works has been interpreted as an obstacle to overcome \cite{xueli,sumit}.
As a possible approach to tackle this issue, in \cite{xueli}, authors have proposed beam switching from a line-of-sight (LOS) link to a collection of non-LOS links when the LOS link breakage occurs.
By utilizing a group of non-LOS links, BSs can bypass the obstacles.
In \cite{sumit}, the multi-hop architecture is designed to compensate for the short LOS distance and to enhance the communication success probability.
However, our study focuses on exploiting the blockage-vulnerable nature of mmW signals to filter network interference.

In addition, it is also worth mentioning that mmW communication is amenable to dense networks since BS densification can assure LOS conditions for more mmW transmissions, thereby relieving the severe distance attenuation problem \cite{Ericsson15,JH15,beam_coord_mag,bwref}.
From this BS densification prospective, we expect that our algorithm will lead to network wide capacity improvements.



\subsection{Contributions and Organizations}
This paper proposes a \emph{building-aware association} scheme to decrease the network interference in dense mmW networks by making BSs adjacent to buildings shape their beam signals toward the buildings.
 The main contributions of this paper are listed as follows:
\begin{itemize}
\item We proposed the building-aware association algorithm that mitigates mmW interference by exploiting the spatial information of buildings. The corresponding $\SIR$ coverage probability and average rate expressions are derived using stochastic geometry.

\item We provided a spatial model of UE locations, capturing the user concentration around buildings. 


\item We optimized the directional association bias $\beta$ of the proposed algorithm. For a special case when the ratio of BS density to UE density is extremely large, i.e. ultra-dense networks \cite{JH15}, we derived the closed-form optimal $\beta$. We thereby provided the design guideline: $\beta$ should properly be increased to reduce more interference, as the density and sizes of buildings increase.

\item We validated the effectiveness of the proposed association algorithm using real building geographic data. Compared to a traditional RSRP-based association, the proposed algorithm is always superior, and achieves higher average rate gains under lower building density locations.

\end{itemize}

The rest of the paper is organized as follows.
The system model is specified in Section II.
The $\SIR$ coverage and average rate in the building-aware association algorithm are derived in Section III.
The results are used to derive the optimal bias $\beta^*$ in Section IV.
Numerical evaluation using real geometric data is presented in Section V.
Finally, the concluding remarks are stated in Section V followed by the proofs of propositions and corollaries in the Appendix.

\section{System Model and Proposed Association Scheme}


\subsection{Network Model}

We consider a dense mmW downlink cellular network where BSs, UEs, and buildings are densely located as in an urban outdoor hotspot.
To reflect the randomness of the urban BS deployment, we assume the BS coordinates $\Phi_b$ follow a homogeneous Poisson Point Process (PPP) with density $\lambda_b$.
We assume the urban buildings form a Boolean model of rectangles with average length $\L$ and width $d_w$, where $\L > \W$. 
The building center coordinates follow a homogeneous PPP $\Phi_\ell$ with density $\lambda_\ell$, independent of $\Phi_b$.

We assume the LOS of each BS is guaranteed when its distance to a typical UE is within an average LOS distance $R_L$.
Such distance is determined by the geography of building blockages, $R_L=\frac{\pi\sqrt{2 \exp\(-\lambda_\ell d_l d_w\)}}{2 \lambda_\ell \(d_l+d_w\)}$ \cite{Bai14}.
Note that the deterministic LOS model provides an analytical simplicity, with only a minor gap in terms of $\SIR$ coverage compared with the random shape theory blockage model based result in a dense BS regime \cite{beam_step}.
For simplicity, we neglect the effect of non-LOS signals\footnote{Note that the effect of non-LOS links might be dominant in practice mmW communication due to the reflections \cite{mmw_intro2,Bai14}. However, their channel gains are typically $20$ dB weaker than the LOS channel gains \cite{nloslos}. Besides, in a dense network regime we are interested in, the mmW communication performance is highly limited by the LOS interferers, as verified in \cite{Bai15} through simulation tests. See \cite{Bai15} for the non-LOS link effects.}, thus only BSs in the LOS region can transmit signals to a typical UE.
We further consider that the LOS probabilities of different links are independent by ignoring the case in which a building blocks the links from neighboring BSs simultaneously \cite{Bai15}.
The assumptions used in modeling the LOS links enables the analytical tractability, in return for reducing its accuracy. 
In section~\ref{sec:simul}, we validate that the effects of these assumptions are quite small through the simulation tests.

\subsection{UE Model}
\begin{figure*}
\centering 
   \subfigure[D-BS condition]{\centering
     \includegraphics[width=6cm]{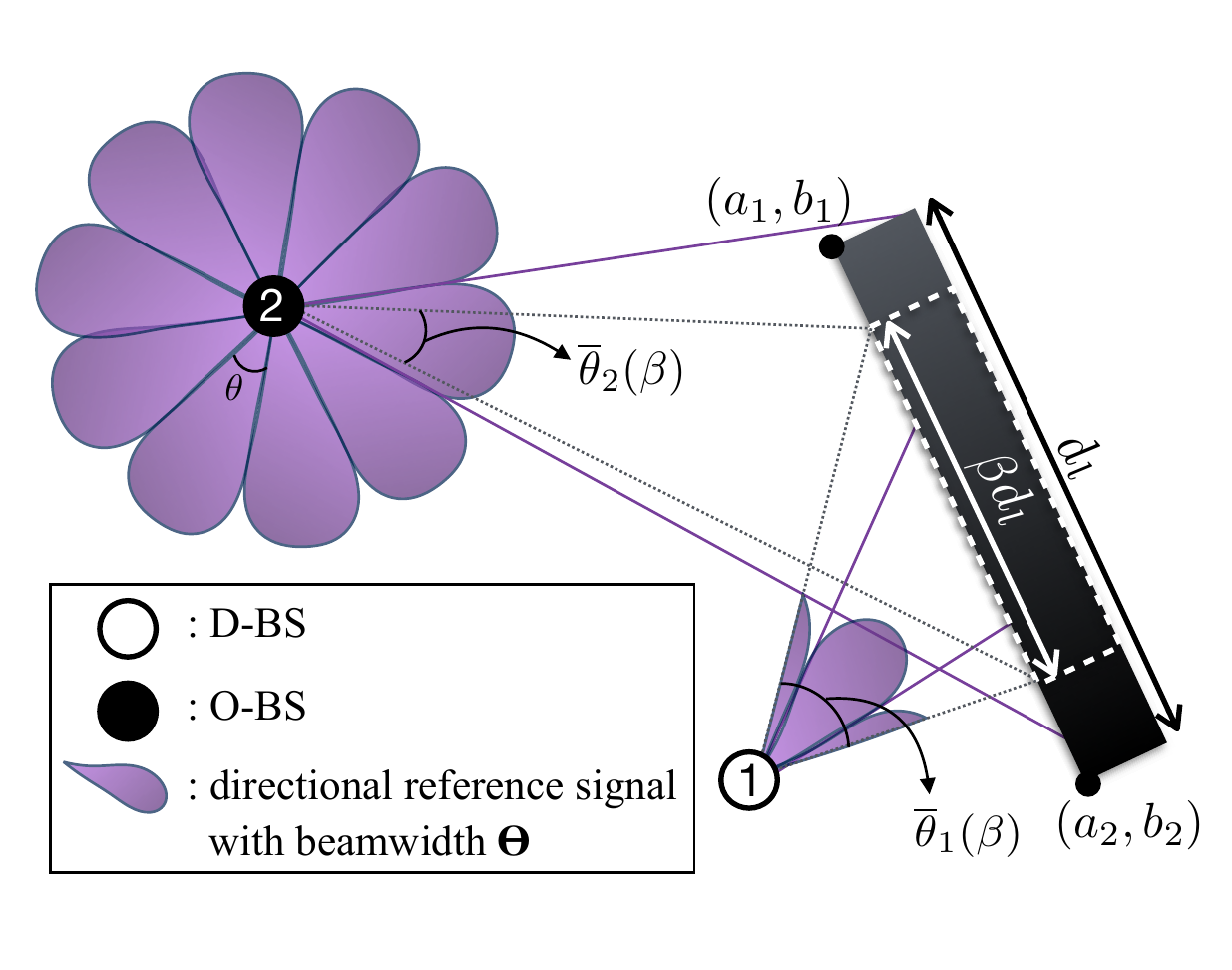} \label{Fig:bldecision_a}} 
   \subfigure[UE association flow chart]{
     \includegraphics[width=7.5cm]{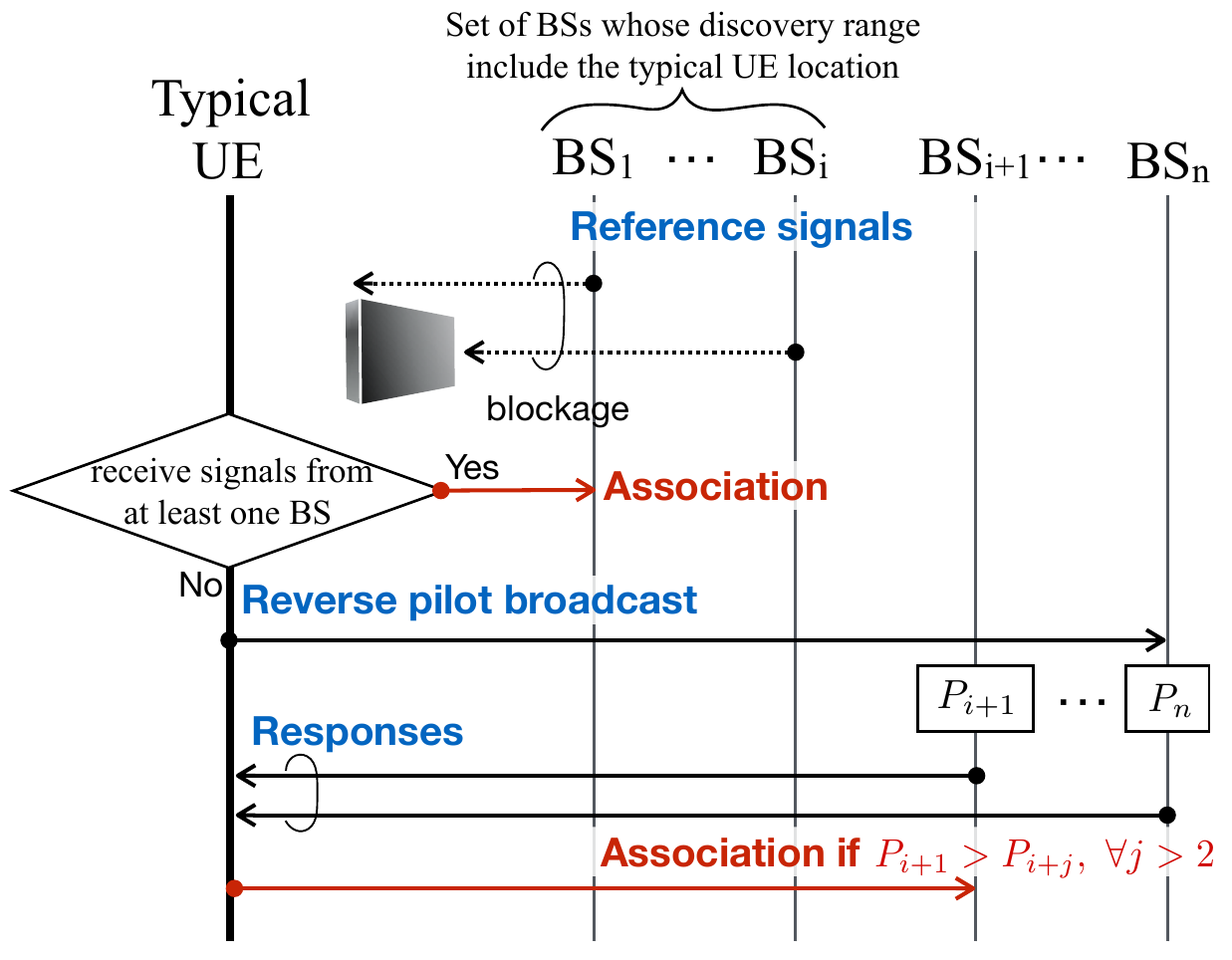} \label{Fig:bldecision_b}} 
     \caption{Illustrations of building-aware association operations: \textbf{D-BS condition.} A BS becomes D-BS if its transmitted signal toward the nearest building does not leak beyond the building whose length is considered as $\beta d_l$, i.e. if $\theta<\bar{\theta}(\beta)$. Then BS 1 transmits the reference signal only toward the building with the discovery range $\bar{\theta}(\beta)$. \textbf{UE association.} Each BS transmits reference signal to its discovery range. When a typical UE receives at least one signal, it associates with the BS who gives the maximum power. If not, it broadcasts a pilot signal to discover the BS. }
\end{figure*}

Consider an urban outdoor area where many UEs are concentrated around the buildings.
We hereafter let $U_n$ denote UEs located near the buildings, e.g., UEs sitting on a bench or a cafe terrace, or traveling along sidewalks.
UEs that are not near the buildings are denoted by $U_r$.
Indoor UEs are beyond the scope of this study since they do not cause interference with outdoor UEs, by considering that indoor and outdoor UEs utilize separate spectrum bands for communications or that even if they utilize the same spectrum, the transmitted signal cannot penetrate walls.

To establish criteria for determining whether UE is close to a building, we assume that each building has an area around it with a width of $d_c$ as shown in Fig.~\ref{system}.
The UE in this area is regarded to be adjacent to the building.
For clarity, let $\mathcal{A}_n$ denote the area around the buildings, and the area excluding $\mathcal{A}_n$ is denoted by $\mathcal{A}_r$.
UE locations $\Phi$ follow a non-homogeneous PPP which are correlated with the building locations $\Phi_\ell$, and the corresponding areas $\mathcal{A}_n$ and $\mathcal{A}_r$.
The local intensity function of $\Phi$ has conditional intensity depending on the spatial location $x$ \cite{inhomo}, $\lambda(x) = \lambda_n \mathbbm{1}_{x \in \mathcal{A}_n} + \lambda_r \mathbbm{1}_{x \in \mathcal{A}_r} $, where $x \in \mathbb{R}^2$.
The UE process $\Phi$ can be interpreted as a mixture of two homogeneous PPPs \cite{mixture}, i.e., when it is located in $\mathcal{A}_n$ it follows PPP with density $\lambda_n$ otherwise it follows PPP with density $\lambda_r$.
The total integrated intensity $\int_{\mathbb{R}^2} \lambda(x) dx$ is denoted by $\lambda$.

To reflect the UE-concentration phenomenon, we let $\gamma_c$ denote the user concentration ratio around buildings.
The ratio is interpreted as i) the average time fraction that a typical UE is adjacent to a building, or ii) the average fraction of UE that is near buildings at a randomly chosen time.
The UE densities $\lambda_n$ and $\lambda_r$ thus vary with the building parameters such as building density $\lambda_\ell$ or length $d_l$, and the user concentration ratio $\gamma_c$.
In this study, we assume that the network can infer this concentration ratio by using UE location statistics such as the spatio-temporal congestion patterns resulting from the daily routines \cite{Klaus03}.

Note that this non-homogeneous UE distribution does not prevent the use of PPP techniques such as Slyvnyak's theorem \cite{kendall} required to analyze $\SIR$, because it does not violate the isotropy in the transmitter point process (i.e. $\Phi_b$).

\subsection{Channel Model}

To form directional beams, all BSs use analog beamforming implemented by utilizing phase shifters.
Each BS steers the antenna direction of the antenna to achieve the maximum array gain at the associated UE.
The error in angle estimation is ignored. 
For analytical tractability, the actual array pattern is approximated by a step function that quantizes the antenna array gains in binary form \cite{Bai15,analogbeam, beam_step}.
In this model, we assume the array gain within the half-power beamwidth $\theta$ radian is identical to the maximum power gain, denoted by $g_m$.
The side lobe gain is denoted by $g_s$.

Each BS transmits a signal with unity power by utilizing the same frequency whose bandwidth is denoted by $W$.
The transmitted signal experiences path-loss attenuation with the exponent $\alpha > 2$ and Rayleigh fading\footnote{
Due to the blockage-vulnerability of mmW links, it is ideal to utilize different path loss exponents to the LOS and non-LOS links \cite{beam_step}. 
Since the LOS links are dominant to the mmW communication in a dense BS regime, we ignore non-LOS links as mentioned above.
For the LOS links, the path loss exponent should be set close to $2$ as discovered from channel measurement results \cite{mmw_intro1, pathloss}.
In addition, although the Rayleigh model may not fit well with the real mmW environment due to the LOS dependent mmW signals, this model can simplify the analytical expressions while providing a lower bound of the downlink rate under Nakagami fading, as mentioned in \cite{JH15,Elshaer16,Li17}. Simulation results have proved that using a Nakagami fading does not provide any additional design insights \cite{Gupta16}. Furthermore, measurement shows that small-scale fading has a relatively little influence on the mmW communications \cite{mmw_intro1}, more relieving our concern to use the Rayleigh model.} with unity mean, i.e. $h \sim \exp(1)$.

Hereafter, we only consider the network in an interference-limited regime, i.e. we neglect the noise power.
Note that a mmW signal is more affected by noise power due to its wide communication bandwidth.
Noise power elimination in this study, nevertheless, provides better tractability while being highly in accordance with the urban dense network scenario we are interested in.
The effect of noise is numerically validated to be negligibly small in Section~\ref{sec:simul}.

Remember that our building-aware association scheme makes BSs adjacent to the buildings transmit signals only toward the buildings by shrinking their UE discovery ranges.
Thus, a BS's signal interferes with a typical UE located at the origin $o$ when the following two conditions hold: i) the distance to the typical UE is within the LOS distance $R_L$, and ii) the origin $o$ is included in its discovery range.
By using Slyvnyak's theorem \cite{kendall}, the $\SIR$ at a typical UE is
\begin{equation}
\small\SIR:=\frac{\mathbbm{1}_{R_L}(|x_0|) g_m h {|x_0|}^{-\alpha} }{ \sum_{x_i \in \Phi_{{b}\setminus x_0} } { \mathbbm{1}_{R_L}(|x_i|) G_i h_i  {|x_i|}^{-\alpha}  }}
\end{equation}\normalsize\noindent
where $\mathbbm{1}_{R_L}(r)$ is an indicator function which returns unity if $r \leq R_L$, $x_0$ and $x_i$ for $i \in \{1,2,\cdots\}$ represent the association BS coordinates and the $i$-th nearest interfering BS coordinates, respectively.
The notation $|\cdot|$ indicates the Euclidian norm.
$G_i$ means the directivity gain in the link from the $i$-th BS.
The UE association rule is specified in the following section.
Considering multiple associations at a BS, the BS selects a single UE per unit time slot using a uniformly random scheduler \cite{tse}.

\subsection{Building-aware UE Association}

Within the building-aware association framework, every BS periodically transfers directional reference signals for UE association \cite{initial}, but the discovery ranges of the BSs are not identical.
We thus decompose the BSs into two types: BSs transmitting the reference signals in every direction (O-BSs); and BSs transmitting the signals only in the direction of a building (D-BSs).
The specific building-aware association scheme follows the rules stated below:
\begin{enumerate}[$\hspace{10pt}1.$]
\item{\textbf{BS: Discovery Range Decision.} Assume every BS knows its location and the nearest building wall's two vertices locations (see ($a_1,b_1$) and ($a_2,b_2$) in Fig.~\ref{Fig:bldecision_a}).
A BS becomes D-BS if its transmitted signal toward the nearest building does not leak beyond the building, whose length is considered as $\beta\L$ as shown in Fig.~\ref{Fig:bldecision_a}.
Then the condition of a BS located at $(x,y)$ is expressed as:
\begin{align}\nn
\theta \leq \overline{\theta}(\beta) = &\left| \text{atan} \[\frac{(1-\beta)b_2+(1+\beta)b_1-2y}{(1-\beta)a_2+(1+\beta)a_1-2x}\] \right.\\
&\left. -\text{atan} \[\frac{(1-\beta)b_1+(1+\beta)b_2-2y}{(1-\beta)a_1+(1+\beta)a_2-2x}\] \right|. \label{condition}
\end{align}\normalsize
The angle $\bar{\theta}(\beta)$ is utilized as the discovery range if the BS satisfies the above condition. Otherwise, its discovery range is $2\pi$.
As long as system parameters such as building density or beam width remain the same, the classification of a BS as a O-BS or a D-BS is maintained once it is determined.
}
\item{\textbf{BS-UE Association}
\begin{enumerate}
\item{Reference Signal Transmission: Each BS transmits reference signals within the discovery range.
When the transmitted signals meet a blockage, their signal strengths highly decrease before reaching to UE (see Fig.~\ref{Fig:bldecision_b}).
}
\item{BS Selection via Signal Strength Comparison: UE listens for reference signals from BSs and transmits a connection request message to the BS that gives the maximum received power. 
}
\item{Reverse Pilot Broadcast:
If the UE detects a radio link failure or does not receive any reference signal, it sends a reverse pilot signal.
BSs that receives the signal transmit their received power strengths to the UE, and the UE connects to a BS whose received power is the strongest.
}
\end{enumerate}
}
\end{enumerate}
The procedure $2$ is periodically repeated to support UE handover or cell reselection.
It is notable that the step $2$-c is to prevent our algorithm making coverage holes.\footnote{When too many BSs associate with users in the building side (i.e., too many D-BSs), users far from buildings are likely to fail to find an associate BS. This might intensify the coverage hole problem in mmW system. To prevent this, we utilize the UE-centric association step 2-c.}
This step resembles the Random Access (RA) procedure in Radio Resource Control (RRC) connection re-establishment in LTE \cite{rre}.
Although this may incur additional control plane congestion, such case rarely happens especially in a dense network topology.


If $\beta=0$, no BS satisfies the condition \eqref{condition} and every BS transmits reference signals to UEs in every direction as in the RSRP-based association scheme.
As $\beta$ increases, the condition gives a loose constraint, increasing the number of D-BSs.
At last when $\beta$ becomes $1$, all the BSs whose transmitted signals toward the nearest buildings do not leak beyond that building become a D-BS.
The optimal bias for maximizing the average data rate is provided in the rest of this paper.
It is worth mentioning that such association bias $\beta$ is identically applied to all BSs even if the surrounding building characteristics of each BS are different.
By so doing, the proposed algorithm obtains a one-shot solution and reduces the association decision time, yet in return for giving up finding the global optimal solution.

Note that with our algorithm, UE may regard that the building number is increased.
This is because the buildings make their adjacent BSs transmit signals only to that building, so that they do not transmit main lobe signal to the UE on the opposite side of the building.
From the UE viewpoint, it becomes as if the building obstacles increase.
Thus in this study, we define the average distance from BS that can transmit main lobe signal as $R_\beta$ (i.e. $R_\beta<R_L$), whose specific expression is derived in Section III.

\section{Coverage and Rate under Building-Aware Association}

The representation of mmW downlink rate using the building-aware association scheme is of prime concern in this section.
The derived result will play a salient role in finding the rate-maximizing bias $\beta$ in Section IV.
In our building-aware association algorithm, a UE receives a distinct data rate according to several random variables such as signal distance, interfering BS locations, building locations, and cell coverage.
The rate in this section thus represents the average data rate of a randomly picked UE considering these random variables.

\subsection{$\SIR$ Coverage} \label{sec:sircov}
We preliminarily represent the $\SIR$ coverage, defined as $\mathsf{P}\(\SIR>t\)$, in terms of the building-aware association bias $\beta$.
To derive the $\SIR$ coverage, we utilize two approximations that are feasible under an urban dense mmW network regime: (i) \emph{homogeneous main beam interferer thinning} with the thinning probability $\frac{\theta}{2\pi}$; and (ii) the \emph{average LOS region}.
Simulation results in Section~\ref{sec:simul} verifies the validity of each.

 \begin{figure*}[b]
\hrulefill
\medmuskip=0mu\thinmuskip=0mu\thickmuskip=0mu 
\footnotesize\begin{align} 
\mathcal{S}_n(\beta)   &= F_0^{R_1}\[ 1+ \ln \(\frac{1+r_2}{1+t^{-1}}\)^{P_\ell t} +  \ln\(\frac{1 + r_1}{1 + r_2}\)^{p_a t} + \ln \(\frac{1+\frac{{R_L}^2}{r^2 t}}{1+r_1}\) ^{\frac{g_s t}{g_m}}\] + F_{R_1}^{R_L}\[ 1+ \ln\(\frac{1+r_1}{1+t^{-1}}\)^{p_a t}+ \ln \(\frac{1+\frac{{R_L}^2}{r^2 t}}{1+r_1}\) ^{\frac{g_s t}{g_m}}\] \label{Eq:approxSIR_c} \\
\mathcal{S}_r(\beta)  &=2F_0^{R_\beta}\[2+\ln \(\frac{t+{R_\beta}^2 r^{-2}}{1+t}\)^{2 p_a t }+\ln \(\frac{t+{R_L}^2 r^{-2}}{t+{R_\beta}^2 r^{-2}}\)^{ \frac{2 g_s t}{g_m}  }\]+2F_{R_\beta}^{R_L} \[2+ \ln \(\frac{t+{R_L}^2 r^{-2}}{t+1}\)^{ \frac{2 g_s t}{g_m}  }\] \label{Eq:approxSIR_n}\\
\text{where}&\quad F_a^b(x) = \int_{a}^b \pi \lambda_b r \exp(-\frac{\pi}{2} \lambda_b r^2 x)dr,\quad p_a=\frac{\theta}{2\pi} +\frac{2\pi-\theta}{2\pi} {\(\frac{g_s}{g_m}\)}^{\frac{2}{\alpha}}, \quad P_\ell=\(\frac{(\pi-\theta)^2}{4 \sin^2(\theta)}+\frac{1}{4 \tan (\theta)}\)\frac{8 \tan^2\(\frac{\theta}{2}\)}{\pi}\(1-p_a\)+p_a,\\
&\quad R_1=\min\(R_L-R_\beta,\frac{R_L}{2}\), \quad r_1 = {\max\[\max\(R_\beta,\frac{R_L}{2}\)^2,r^2\]}{\(r t^{\frac{1}{\alpha}}\)^{-2}}, \quad \text{and}\quad r_2~=~{\min\(R_L-R_\beta,\frac{R_L}{2}\)^2}{\(r t^{\frac{1}{\alpha}}\)^{-2}}
\end{align}\normalsize
\end{figure*}

\begin{figure}
\centering 
{\includegraphics[width=8cm]{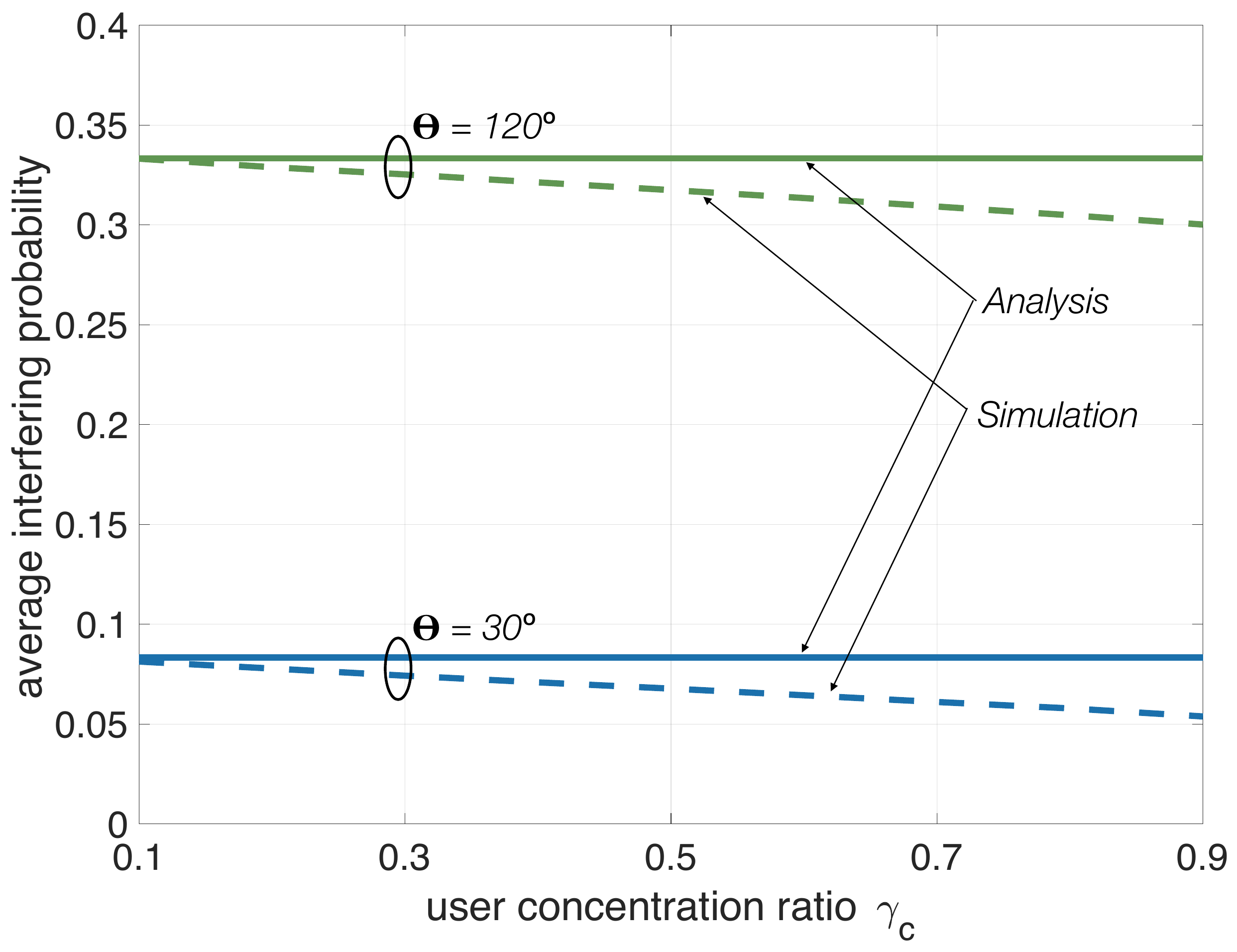}}
\caption{Average main-lobe interfering probability of BSs according to the user concentration ratio $\gamma_c$ and beam width $\theta$ ($\lambda= \text{400 BSs}/\text{km}^2$, $\lambda_\ell=\text{400 buildings}/\text{km}^2$, $d_l=30$ m, $d_w=10$ m, $t=15$ dB).\label{Fig:veri_prob}} 
\end{figure}

The first approximation tackles the non-homogeneity of the interfering BSs with the main lobe gain $g_m$.
This mainly results from the random UE selections in BSs which depend on the coverage areas of BSs and the non-uniformly distributed UE locations.
In a dense network, however, the cell coverage area tends to be equal, diminishing the cell area dependency.
Although this approximation gets loose as the non-homogeneity of the UE distribution is severe, we validate that the average main beam interfering probability of BSs is  consistent with the thinning probability $\frac{\theta}{2\pi}$ when the concentration ratio is high through simulation experiments (see Fig.~\ref{Fig:veri_prob}).

\begin{figure}     
\centering
   \subfigure[For $U_r$]{\centering
     \includegraphics[width=4cm]{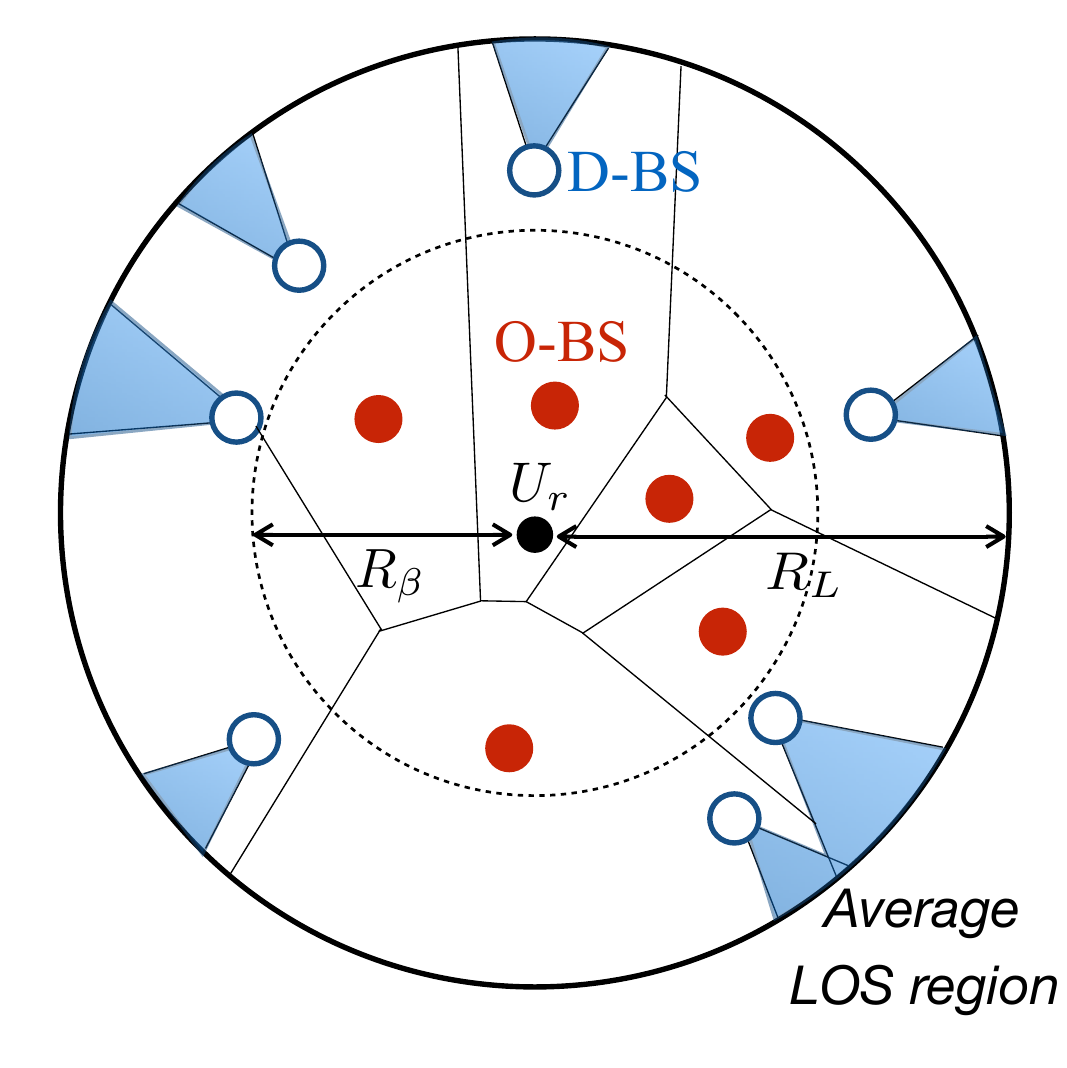} \label{rbeta_ur}}
   \subfigure[For $U_n$]{
     \includegraphics[width=4cm]{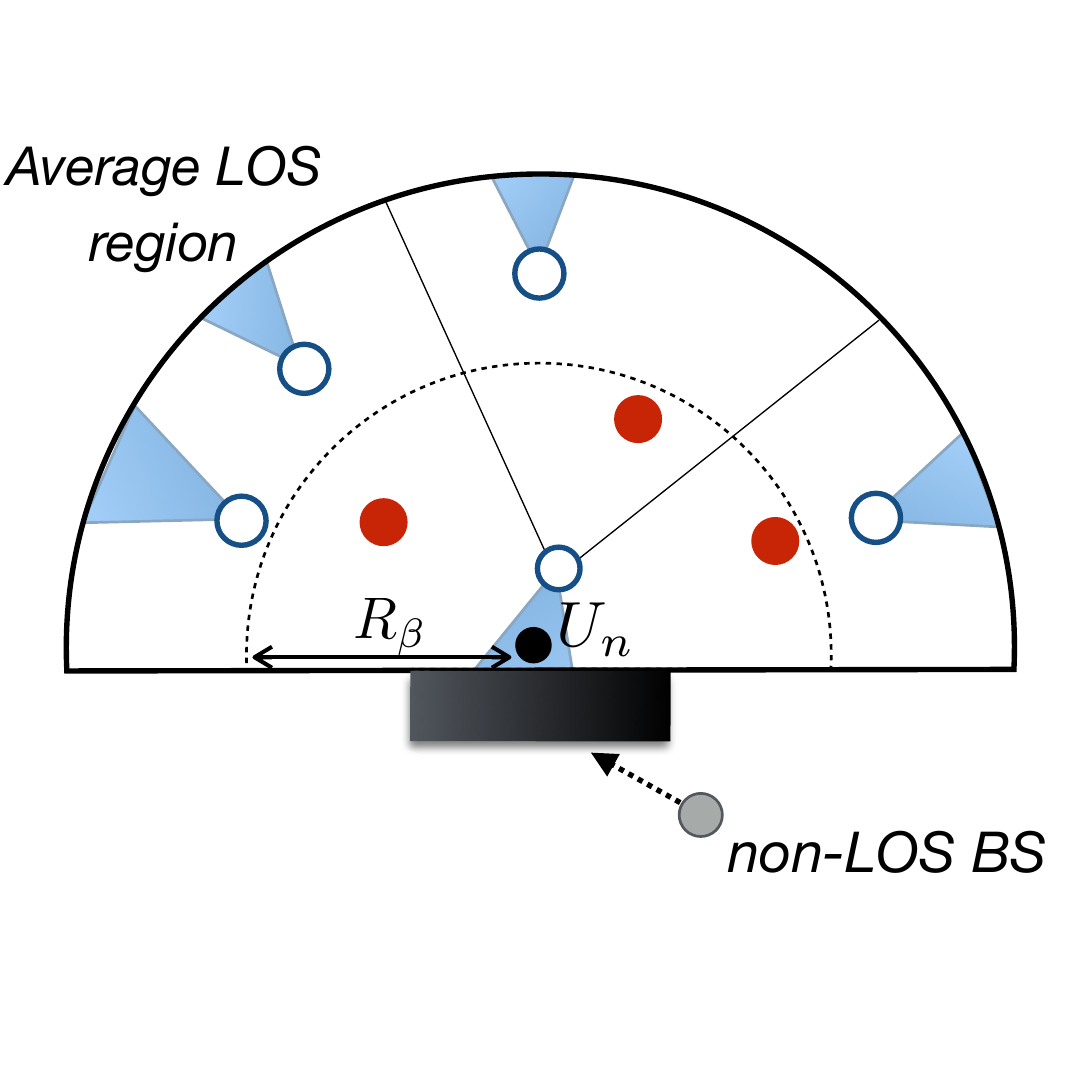}\label{rbeta_un}} 
\caption{Illustrations of the average LOS region and region of BSs that can interfere to a typical UE with a main lobe gain $g_m$. } \label{rbeta} 
\end{figure}

The second approximation is motivated by the average LOS ball model in \cite{Bai14}.
From the perspective of a typical $U_r$, which is remote from the buildings, the LOS region is approximated to a ball area within an LOS distance $R_L$.
On the other hand, from the perspective of a typical $U_n$, which is near a building, BSs located across the building cannot transmit any signals.
Thus the average $U_n$ LOS region becomes a half moon shape with a radius $R_L$ as shown in Fig.~\ref{rbeta_un}.
Hereafter we assume that a typical $U_n$ is located at the center of the building wall and its distance to the building is negligible.
This average LOS region model may reduce the analytical accuracy.
However, since we consider the urban scenario where buildings are densely located, the difference between the actual and approximated LOS regions decreases, relieving this concern.

In the LOS region, D-BSs transmit signals toward the buildings, i.e. in the opposite direction of the typical UE located at the center of the LOS region (see Fig.~\ref{rbeta}).
Such D-BSs transmit interfering signals only with the side lobe gain $g_s$.
Thus, we define the average region of BSs that can transmit main lobe signals as follows, by considering the farthest distance from the building to the D-BS.

\begin{lemma}\emph{In our algorithm, the maximum distance of D-BS from the building is $\frac{\beta d_l}{2 \tan \(\frac{\theta}{2}\)}$. Therefore, the region of BSs that can transmit main lobe signals to a typical UE has a radius of $R_\beta=R_L-\frac{\beta d_l}{2 \tan \(\frac{\theta}{2}\)}$. 
\vskip 0pt \noindent
{\bf  Proof.} 
See Appendix.
 \hfill $\blacksquare$}
\end{lemma}

This result implies that for a typical $U_n$, distance $R_\beta$ cannot be less than $\frac{R_L}{2}$.
This is because, from the perspective of $U_n$, within the maximum distance of $\frac{\beta d_l}{2 \tan \(\frac{\theta}{2}\)}$, there are D-BSs that transmit signals toward the $U_n$'s nearest building and interfere with the $U_n$ (see Fig.~\ref{rbeta_un}).
Note that if a BS satisfies the D-BS condition for more than one building, the BS only transmits signals to the nearest building.
That is, when a BS is closer than the maximum distance $\frac{\beta d_l}{2 \tan \(\frac{\theta}{2}\)}$ not only from the building where the typical $U_n$ is attached but also from the building at the edge of the average LOS region, i.e., $R_\beta < \frac{R_L}{2}$, the BS chooses the nearest building to transmit signals.
Thus, the modified $U_n$ LOS region no longer depends on $\beta$ after the bias $\beta$ increases to make $R_\beta=\frac{R_L}{2}$.

By considering the path-loss exponent of the LOS links is close to $2$ and utilizing the above approximations and lemma, we can derive the $\SIR$ coverage as in the following proposition.

\begin{proposition}\emph{($\SIR$ Coverage) The $\SIR$ coverage of a typical UE becomes
\begin{small}\begin{align}
\mathcal{S} = \gamma_c \mathcal{S}_n(\beta) + (1- \gamma_c) \mathcal{S}_r(\beta),\label{Eq:SIRcov}
\end{align}\end{small}
\noindent where $\mathcal{S}_n(\beta)$ and $\mathcal{S}_r(\beta)$ represent the $\SIR$ coverage of a typical $U_n$ and $U_r$, respectively. 
They are given as \eqref{Eq:approxSIR_c} and \eqref{Eq:approxSIR_n} at the bottom of the page.
\vskip 0pt \noindent
{\bf  Proof.} 
See Appendix.
 \hfill $\blacksquare$}
\end{proposition}

\begin{figure}
\centering 
{\includegraphics[width=8cm]{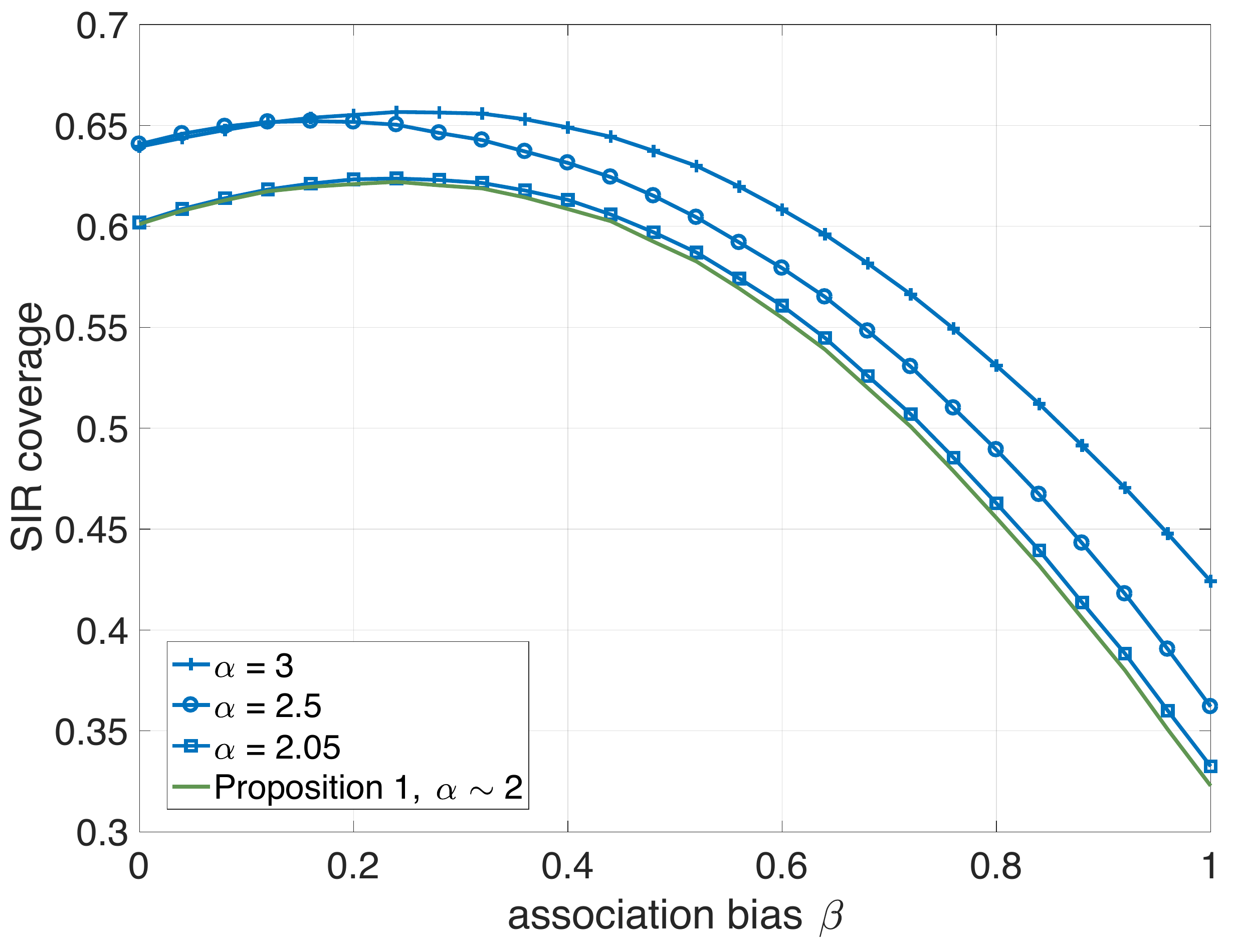}} 
\caption{ {$\SIR$ coverage according to the path-loss $\alpha$ ($\lambda_b= \text{400 BSs}/\text{km}^2$, $\lambda_\ell=\text{400 buildings}/\text{km}^2$, $\theta=\frac{\pi}{6} rad$, $d_l=30$ m, $d_w=10$ m, $t=10$ dB, $\gamma_c=0.6$, ${g_m}=20$ dB, ${g_s}=0$ dB)}.} \label{Fig:alpha}
\end{figure}

Note that one of key enablers to provide the $\SIR$ coverage is the assumption that the path-loss exponent $\alpha$ is equal to $2$ since we mainly focus on the LoS links \cite{pathloss}.
Fig.~\ref{Fig:alpha} shows that the $\SIR$ coverage according to different $\alpha$, verifying that the $\SIR$ coverage gap is small when $\alpha$ is larger than $2$. 
Although such gap increases with $\alpha$, they have similar tendencies.

The result shows that the association distance of a typical UE does not change with the bias $\beta$.
This follows from the average LOS ball model in which the typical UE is located at the center of the LOS region (see Fig.~\ref{rbeta}).
When the typical UE is $U_r$, its associated BS is located farthest from the buildings within the LOS ball region, i.e. it does not become a D-BS, maintaining the signal distance.
When the typical UE is $U_n$, even if the associated BS becomes a D-BS, the connection is not cut off because the signal is transmitted toward the building to which the UE is attached.
However, in actual networks the proposed algorithm might lengthen the UE association distances.
This implies that the result in Proposition 1 deals with the upper-bound case from the signal strength perspective.

Note that the $\SIR$ coverages $\mathcal{S}_n(\beta) $ and $\mathcal{S}_r(\beta) $ have distinct tendencies according to the bias $\beta$.
The following remark specifies these behaviors by considering two extreme cases.

\begin{remark} If $\gamma_c=0$, i.e. when UE is far from buildings at all times, the $\SIR$ coverage monotonically increases with $\beta$. If $\gamma_c=1$, i.e. when UE is always near a building, the $\SIR$ coverage is convex-shaped over $\beta$.
\end{remark}
This is mainly because, from the $U_n$ perspective, a higher $\beta$ makes more BSs transmit signals toward the buildings.
In other words, high $\beta$ decreases the LOS region (or $R_\beta$) while maintaining the signal strength.
So the interference decreases.
However, from the $U_c$ perspective, the $\beta$ increment firstly makes BSs located near buildings transmit signals in opposite directions from $U_n$ and reduces its interference.
When $\beta$ is large, even BSs adjacent to $U_n$ may become D-BSs and transmit signals toward the building to which $U_n$ is attached. This incurs more interference to $U_n$.
This finding helps not only to capture the impact of $\beta$ in a more intuitive way, but also to find the optimal $\beta^*$ which will be described in detail in Section IV.

\subsection{Average Rate}\label{sec:rate_analy}
We define the rate of a typical UE as \small$ \mathcal{R} := \mathsf{E}\[\frac{W}{N+1} \]\mathsf{P}\(\SIR>t\) \log\[1+t\],\ $\normalsize 
where $W$ represents the bandwidth and $N$ the UE number at a BS that associates with a typical UE.
The $\SIR$ coverage in the previous section enables the determination of the average spectral efficiency $\mathsf{P}[\SIR>t] \log (1+t)$.
To derive the rate $\mathcal{R} $, in addition, we need to calculate the available amount of bandwidth under the uniformly random scheduler, $\frac{W}{1+N}$.
The random variable $N$, the UE number at a BS associated with a typical UE, can be expressed by deriving the probability density function (PDF) of the cell size \cite{SMYu,offloading,jm}.
In this study, $N$ has to be derived by considering the building geometry since we consider non-homogeneous UE distributions.

To provide a more tractable mathematical representation, we assume UE number in the cell coverage of the BS associated with a typical UE to be its mean value as in \cite{offloading}, resulting in $\mathsf{E}\[\frac{1}{1+N}\] \sim \frac{1}{1+\mathsf{E}[N]}$. 
Although this assumption reduces the analytic accuracy, it does not significantly affect the tendency to rate $\mathcal{R}$ as validated in Section~\ref{sec:simul}.

As mentioned in Section II, the UE density in the area near the buildings is different from that in the area far from the buildings. 
In the following lemma, the corresponding UE densities are given according to the user concentration ratio $\gamma_c$ and the geography of buildings.

\begin{lemma} \label{UE_den}\emph{The UE density $\lambda$ is decomposed into the $U_n$ density, $\lambda_n$, and the $U_r$ density, $\lambda_r$, as follows.
\begin{small}\begin{align}
\lambda_n&= \lambda \gamma_c \[1+\frac{1-2 \lambda_\ell (d_l+d_w) d_c- \lambda_\ell d_ld_w}{2 \lambda_\ell (d_l+d_w) d_c}\],\\
\lambda_r&= \lambda(1-\gamma_c) \[1+\frac{2 \lambda_\ell (d_l+d_w) d_c}{1-2 \lambda_\ell (d_l+d_w) d_c- \lambda_\ell d_ld_w}\].
\end{align}
\end{small}
\vskip 0pt \noindent
{\bf  Proof.} 
A UE is $U_n$ when the distance to its nearest building is less than $d_c$.
Then a unit area can be divided into three parts: i) indoor area has an average size of $\lambda_\ell \L \W$, ii) $U_n$ area, $\mathcal{A}_n$, has an average size of $\lambda_\ell 2 (\L+\W) d_c$, and iii) $U_r$ area, $\mathcal{A}_r$, occupies the rest.
We assume that there is no overlapping area between buildings. so we can derive the densities $\lambda_n$ and $\lambda_r$, respectively.  \hfill $\blacksquare$
}
\end{lemma}

Note that the Lemma \ref{UE_den} enables us to calculate the mean UE number, leading to the following proposition.

\begin{proposition}\emph{(Average Rate) The average rate using the building-aware association scheme is
{\begin{small}
\begin{align}
\mathcal{R}  = \frac{W \gamma_c}{1+ N_n} \mathcal{S}_n(\beta)  \log(1+t) + \frac{W (1-\gamma_c)}{1+N_r} \mathcal{S}_r(\beta)  \log(1+t) , \label{rate}
\end{align}
\end{small}}
\noindent where $N_n$ and $N_r$ represent the mean number of UEs in the same cell coverage with a $U_n$ and $U_r$, respectively, and are given as \eqref{Eq:avgUnum_c} and \eqref{Eq:avgUnum_n} at the bottom of the next page.
\vskip 0pt \noindent
{\bf  Proof.} 
See Appendix.
 \hfill $\blacksquare$}
\end{proposition}

\begin{figure}
\centering 
{\includegraphics[width=8cm]{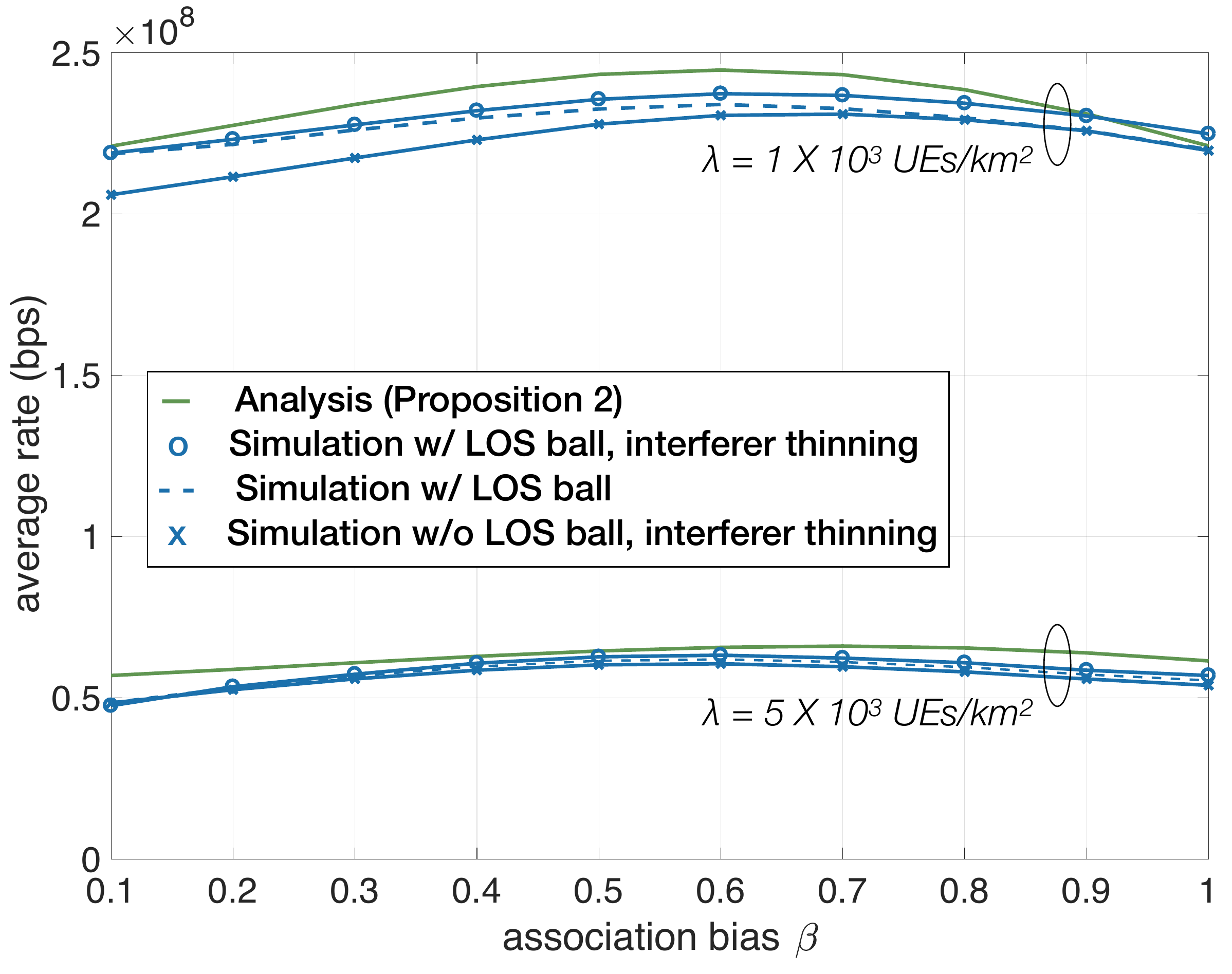}} 
\caption{Comparison of average rate with simulation results ($\lambda_b= \text{200 BSs}/\text{km}^2$, $\lambda_\ell=\text{200 buildings}/\text{km}^2$, $\theta=\frac{\pi}{6} rad$, $d_l=30$ m, $d_w=10$ m, $t=10$ dB, $\gamma_c=0.6$, ${g_m}=20$ dB, ${g_s}=0$ dB). } \label{Fig:simul}
\end{figure}

 \begin{figure*}[b]
\hrulefill

\begin{small}
\begin{align} N_r &=
\frac{1.28 \lambda_r}{\pi\lambda_b {R_\beta}^2} \(\mathbbm{1}_{R_\beta < 0.68 \sqrt{\lambda}}\[\(A_c-A_r\)\lambda_n/\lambda_r  +A_r\] + \mathbbm{1}_{R_\beta \geq 0.68 {\sqrt{\lambda}}}\)
 \label{Eq:avgUnum_n} \\
N_n&=\(E({R_1},{R_L})N_r+0.64\beta\L\[\lambda_n c_1(d_c)  + \( d_c \lambda_n -\frac{d_c \lambda_r}{2}\)E(d_c,R_1) +\lambda_r\(c_1(R_1)-c_1(d_c)\) \]\)/\(1-e^{-\frac{\pi\lambda_b {R_L}^2}{2}}\) \label{Eq:avgUnum_c}\\
\text{where}\;\;\; A_c&=\max\(\pi{R_\beta}^2,\pi {R_L}^2-\frac{\beta \L}{2}\[\pi\lambda_b R_L \({R_L}^2-{R_\beta}^2\)-\frac{2}{3}\pi\lambda_b \({R_L}^3-{R_\beta}^3\)\]\)\quad \text{and} \\ \nn
\quad A_r&=\begin{cases}
\pi\(R_L-d_c\)^2, {\quad\quad \textrm{if $d_c>R_L-R_\beta$}} \\
\max\(\pi{R_\beta}^2,\pi \[R_L-d_c\]^2-\frac{\beta\L\pi\lambda}{2}\[ \(R_L-d_c\)\(\[R_L-d_c\]^2-{R_\beta}^2\)-\frac{2\(\[R_L-d_c\]^3-\[R_\beta-d_c\]^3\)}{3}\]\), {\quad \textrm{otherwise}.} 
\end{cases}
\end{align}\end{small}
The function $E(a,b) = \exp\(-\pi\lambda_b {a}^2/2\)-\exp\(\pi\lambda_b {b}^2/2\)$, $c_1(x)~=~ \erf\(x \sqrt{\pi \lambda/2} \)/\sqrt{2\lambda_b} - x \exp\(-\pi\lambda_b {x}^2/2\)$, and the indicator function $\mathbbm{1}_\mathbb{A}$ returns $1$ if the event $\mathbb{A}$ occurs.
\end{figure*}\normalsize

In our algorithm, the O-BSs have to expand their cell coverages to fill the coverage hole resulting from the coverage reduction of D-BSs (see Fig.~\ref{rbeta}).
In this regard, the event ${R_\beta < 0.68 {\lambda}^{-\frac{1}{2}}}$ in \eqref{Eq:avgUnum_n} represents the case that a BS serving the typical UE has to expand the coverage since its neighboring BSs become D-BSs.

This analytic result is verified through simulation experiments rigorously carried out without approximations used in the analysis, e.g., average LOS region, uniform interfering BS thinning, and mean UE number.
Fig.~\ref{Fig:simul} compares the average rate from the analysis to that of PPP simulation.
Although there is a slight gap between the two results due to the analytic assumptions, they follow a similar trend.
Besides, when including the LOS ball model and homogeneous interferer thinning assumption discussed in Section~\ref{sec:sircov}, the simulation result are more close to the analytic one.
The reason the rate results from the simulations are generally lower than those from the analysis is two-fold.
First, $\mathsf{E}\[\frac{1}{x}\]$ is smaller than $\frac{1}{\mathsf{E}\[x\]}$, which is utilized in the mean UE number approximation in Section~\ref{sec:rate_analy}.
Second, our analysis does not consider the case that the association distance of a typical UE might increase due to the building-aware association, as mentioned above.
In addition, this figure shows that the difference between the optimal bias ${\beta_\mathcal{R}}^*$ from the analysis and the simulation is larger when the UE density is high.
The reason behind is that the mean UE number approximation reduces the analytical accuracy when there are many UEs.

According to Proposition 2, mean UE numbers of the associated cell areas of $U_n$ and $U_r$, i.e. $N_n$ and $N_r$, have remarkable characteristics according to the bias $\beta$ as follows.

\begin{remark} {
The average UE number $N_r$ increases with the bias $\beta$ while $N_n$ decreases with it.
}
\end{remark}
The reason behind is that, when a typical UE is $U_r$, its associated BS is the farthest BS from the buildings in the LOS ball.
Therefore as $\beta$ increases, more neighboring BSs become D-BSs, increasing the typical cell coverage as well as the associated UE number.
When a typical UE is $U_n$, probability that the associated BS becomes D-BS increases with $\beta$, resulting in the associated UE number decrement.

The remark also implies that the $\beta$ increment decreases the amount of available bandwidth of a $U_r$, and vice versa for a $U_n$. 
Remember that higher $\beta$ increases the $\SIR$ coverage of a $U_r$ as explained in Remark 1.
That is, $\beta$ affects the available bandwidth and $\SIR$ coverage in opposite ways, leading us to derive the optimal $\beta$ maximizing the average rate.

\section{{Rate Optimal Building-Aware Association}}

This section deals with the rate-maximizing bias $\beta$ for the building-aware association algorithm.
The optimal $\beta$ derivation is not straightforward because our algorithm affects UE differently depending on whether the UE is close to a building or not.
Furthermore, the bias $\beta$ has the opposite effect on the $\SIR$ coverage and the amount of available bandwidth in the data rate of a typical UE.

Due to the additional UE number consideration, the optimal bias for maximizing the rate differs from that for maximizing the $\SIR$ coverage.
In this section, we first focus on the optimal bias maximizing the $\SIR$ coverage, then provide the rate maximizing $\beta$.

As mentioned in Section III-A, the $\SIR$ coverage $\mathcal{S}_n(\beta)$ is maintained after $\beta={ \tan \(\frac{\theta}{2}\)}\frac{R_L}{\L}$ that makes $R_\beta=\frac{R_L}{2}$ , and $\mathcal{S}_r(\beta)$ increases with $\beta$.
It implies that the $\SIR$ coverage $\mathcal{S}$ is non-decreasing with $\beta$ after ${ \beta=\tan \(\frac{\theta}{2}\)}\frac{R_L}{\L}$.
This characteristic highly reduces the search range of $\beta$, leading to the following Proposition. 

\begin{figure}
\centering 
{\includegraphics[width=8cm]{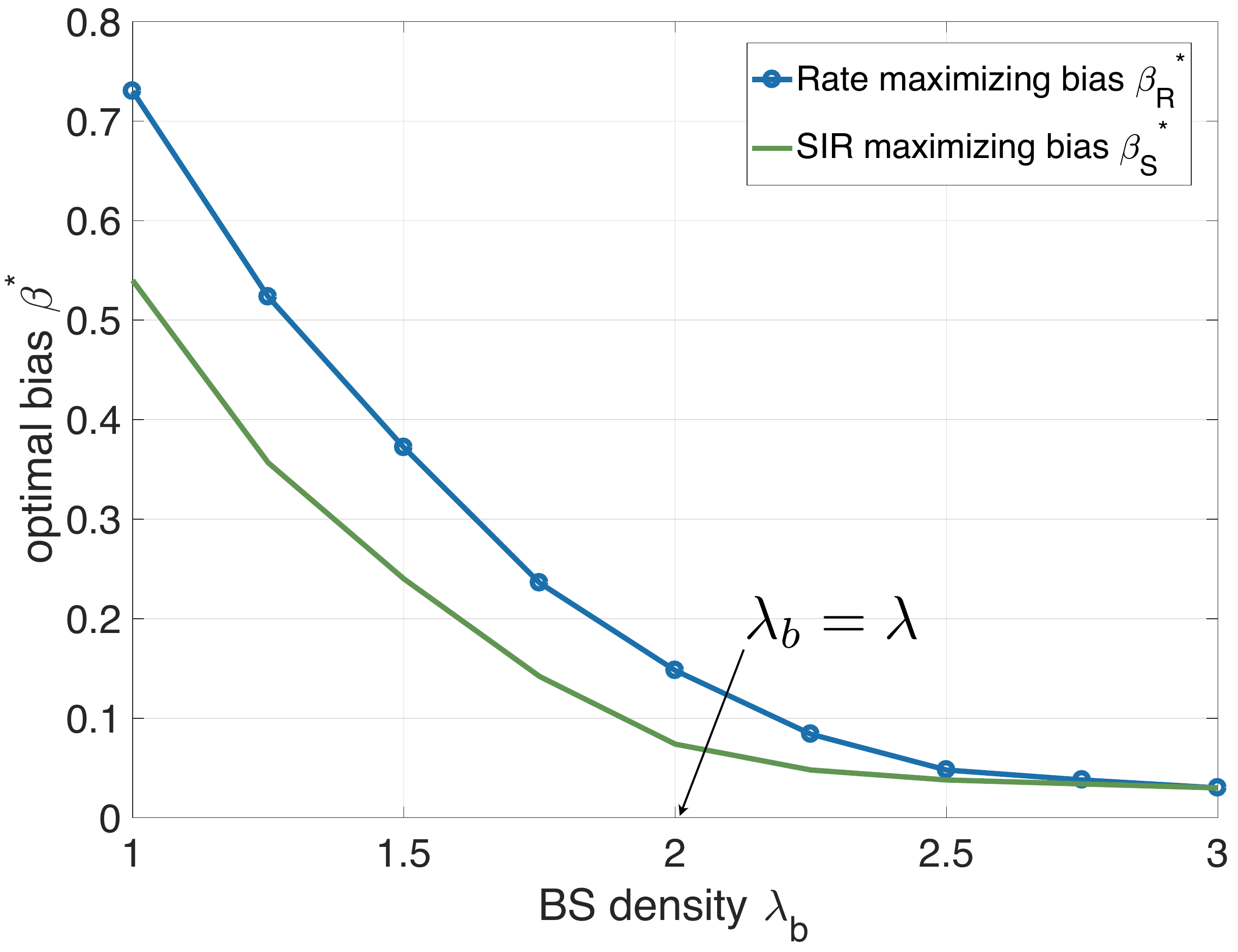}} 
\caption{ $\SIR$ coverage and rate of the building-aware algorithm according to the association bias $\beta$ ($\lambda_\ell=\text{300 buildings}/\text{km}^2$, $\lambda= \text{2 $\times 10^4$ UEs}/\text{km}^2$, $\theta=\frac{\pi}{4} rad$, $d_l=30$ m, $d_w=10$ m, $t=10$ dB, $\gamma_c=0.6$, ${g_m}=20$ dB, ${g_s}=0$ dB).} \label{Fig:bias}
\end{figure}

\begin{figure*}     
\centering
   \subfigure[$\SIR$ coverage]{\centering 
     \includegraphics[width=8cm]{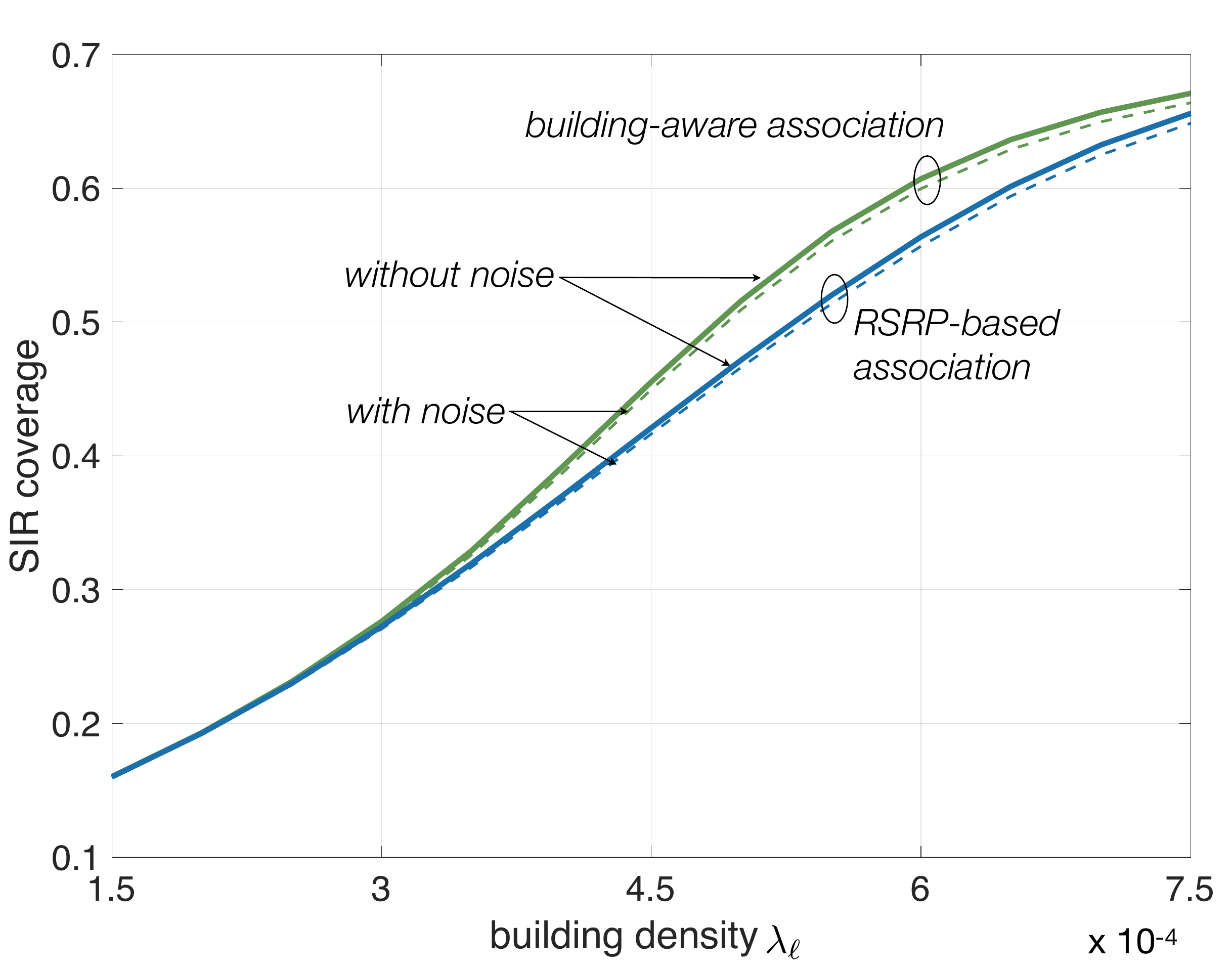} }
   \subfigure[average rate]{
     \includegraphics[width=8cm]{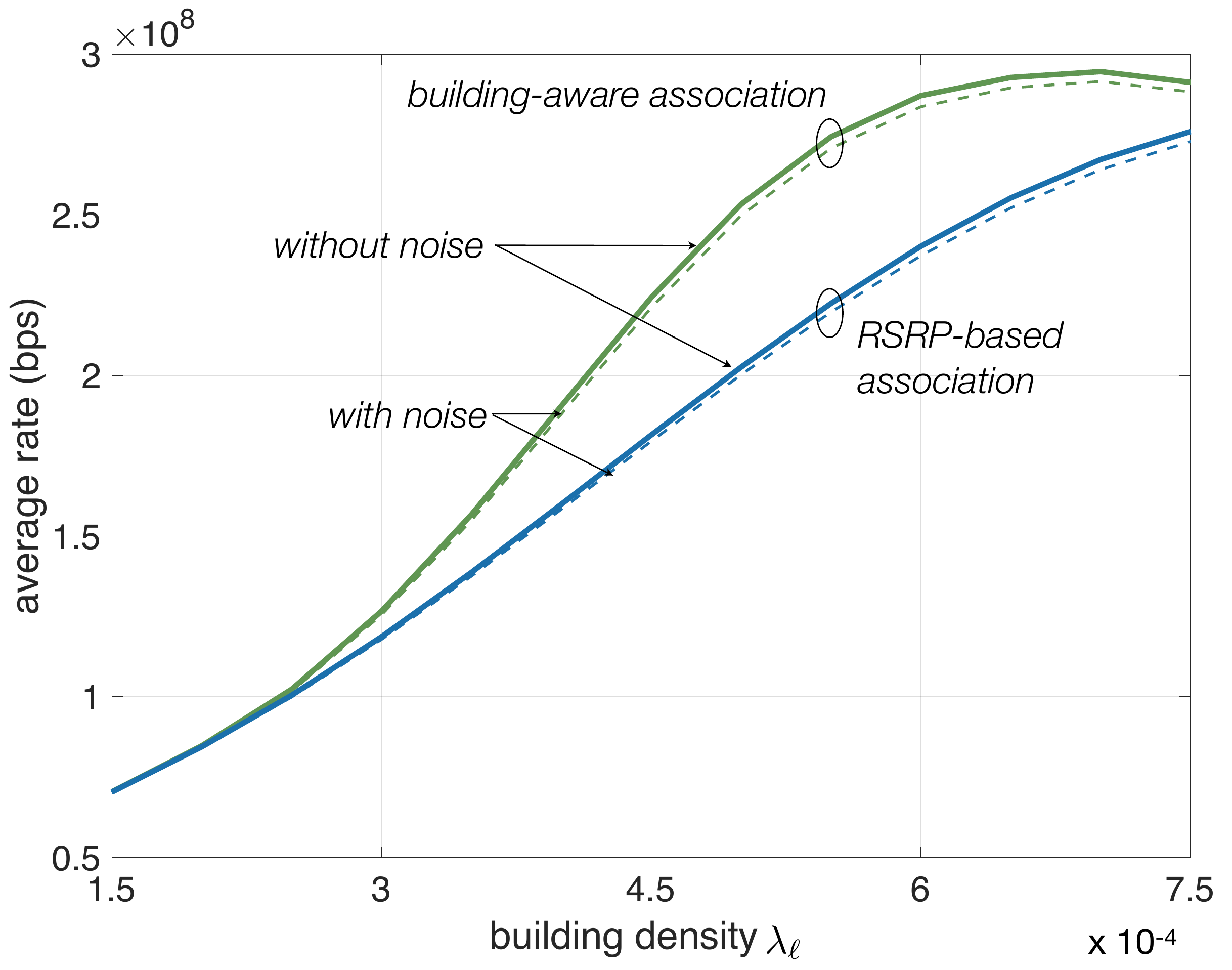}}
\caption{Impact of building density $\lambda_\ell$ on $\SIR$ coverage and rate ($\lambda_b= \text{600 BSs}/\text{km}^2$, $\lambda= \text{2 $\times 10^3$ UEs}/\text{km}^2$, $\theta=\frac{\pi}{4}$, $\gamma_c=0.6$).}\label{Fig:build_den}
\end{figure*}

\begin{figure*}     
\centering
   \subfigure[$\SIR$ coverage]{\centering 
     \includegraphics[width=8cm]{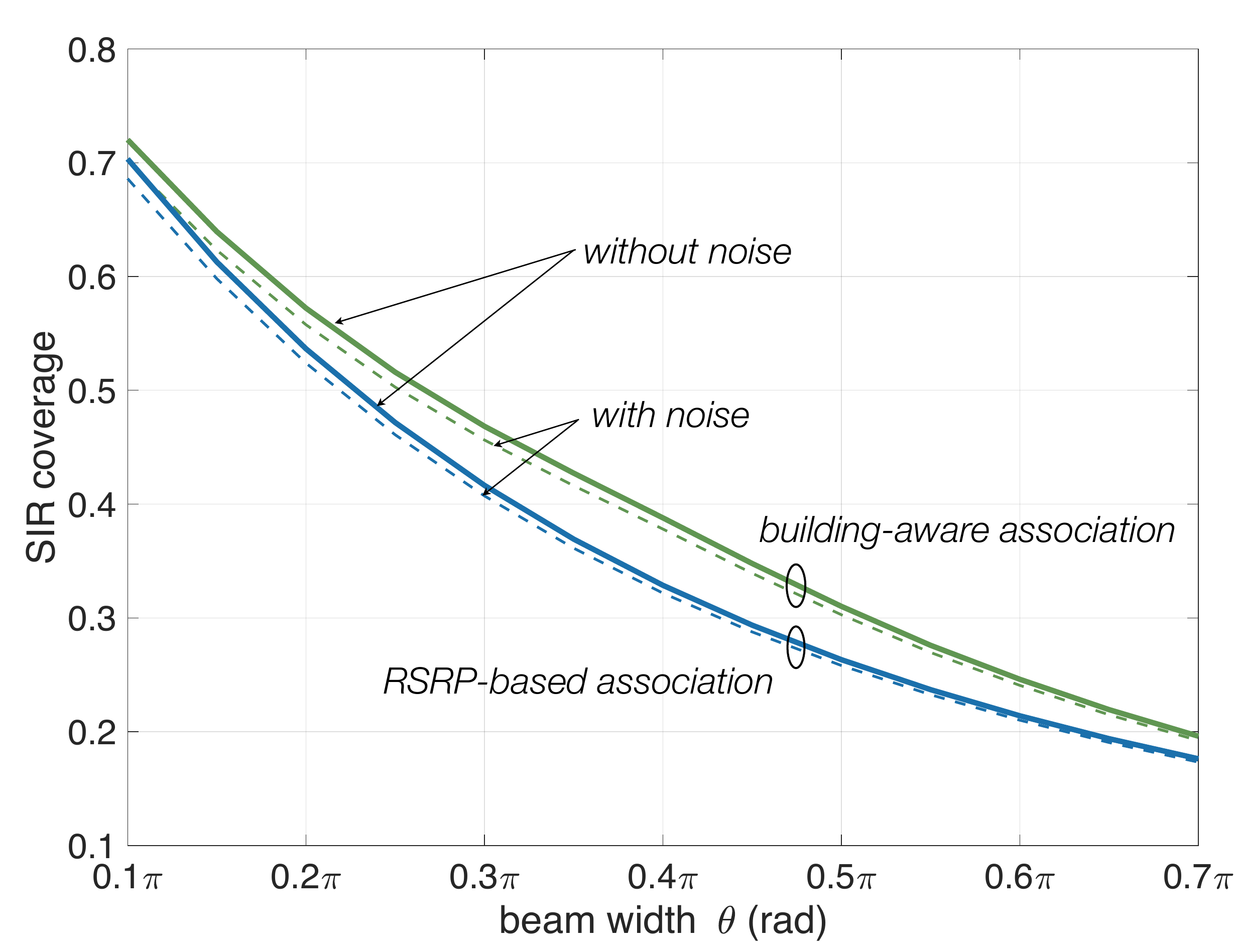} }
   \subfigure[average rate]{
     \includegraphics[width=8cm]{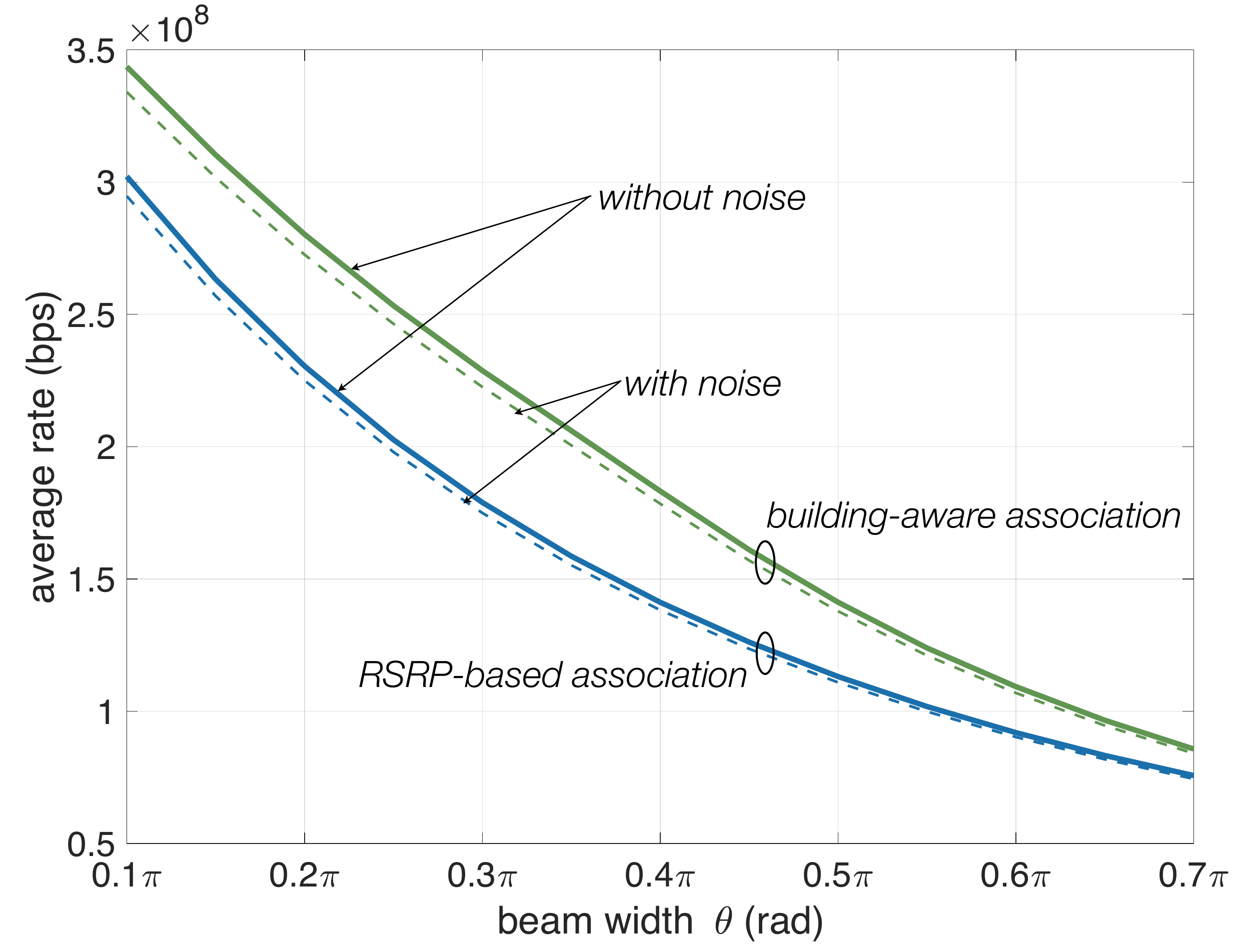}}
\caption{Impact of beamwidth $\theta$ on $\SIR$ coverage and rate ($\lambda_b= \text{600 BSs}/\text{km}^2$, $\lambda= \text{2 $\times 10^3$ UEs}/\text{km}^2$, $\lambda_\ell= \text{$500$ buildings}/\text{km}^2$, $\gamma_c=0.6$).} \label{Fig:beam}
\end{figure*}

\begin{proposition}\emph{(Optimal $\beta$) The optimal association bias is given as below.
\begin{itemize}
\item{For maximizing the $\SIR$ coverage, 
\begin{small}\begin{align}
\beta^*_\mathcal{S} = 
\begin{cases}
\mathbbm{1}_{C_1}+\(1-\mathbbm{1}_{C_1}\)\underset{T_\beta}{\mathrm{argmax}} \ \mathcal{S}  \quad\quad &{\textrm{if ${R_L \tan \(\frac{\theta}{2}\)} < \L$}} \\ \underset{0\leq\beta\leq 1}{\mathrm{argmax}}  \ \mathcal{S}  \quad\quad &{\textrm{otherwise}}
\end{cases}
\end{align} \end{small}
where \small $T_\beta=0 \leq \beta \leq {{ \tan \(\frac{\theta}{2}\)}R_L}{\L}^{-1}$, ${C_1 := {\gamma_c \mathcal{S}_n\(\tan \(\frac{\theta}{2}\)\frac{R_L }{\L}\) + (1-\gamma_c) \mathcal{S}_r(1) >\underset{T_\beta}{\mathrm{max}}  \ \mathcal{S} }}$.  \normalsize
}
\item{For maximizing the average rate,
\begin{small}\begin{align}
{\beta_{\mathcal{R}}}^*=\underset{\beta}{\textrm{argmax}}\   \frac{\gamma_c}{1+ N_n} \mathcal{S}_n(\beta) + \frac{(1-\gamma_c)}{1+N_r} \mathcal{S}_r(\beta) . \label{optbeta_r}
\end{align}\end{small}}
\end{itemize}
\vskip 0pt \noindent
{\bf  Proof.} 
See Appendix.
 \hfill $\blacksquare$}
\end{proposition}

The result indicates that ${\beta_{\mathcal{S}}}^*=1$ if $\gamma_c=0$ since $\mathcal{S}_r(\beta) $ monotonically increases with $\beta$.
Also, it reveals that the optimal bias ${\beta_{\mathcal{S}}}^*$ decreases with the concentration ratio $\gamma_c$, because the $\SIR$ increment of a typical $U_n$ is smaller than that of a typical $U_r$.
Note that ${\beta_{\mathcal{R}}}^*\neq 1$ when $\gamma_c=0$ due to the average UE number $N_r$ increment along with $\beta$.

Although it is difficult to derive the optimal bias ${\beta_{\mathcal{R}}}^*$ in a closed form, it can be determined through a linear search using the above Proposition.
Furthermore, we can obtain ${\beta_{\mathcal{R}}}^*$ more easily in the asymptotic case where the BS density is much higher than the UE density, as shown in the following Corollary.

\begin{corollary}\label{asym_prop}\emph{(Optimal Bias in Ultra-dense Scenario) When $\lambda_b \gg \lambda$, the mean UE number $N_n$ and $N_r$ converge to $0$, i.e. $\lim_{\lambda / \lambda_b\to 0} N_n = 0$ and $\lim_{\lambda / \lambda_b\to 0} N_r = 0$.
Thus, the optimal $\beta$ for maximizing the $\SIR$ coverage and the rate become identical, i.e. ${\beta_{\mathcal{R}}}^* \sim {\beta_{\mathcal{S}}}^*$.
\vskip 0pt \noindent}
\end{corollary}

\begin{figure*}     
\centering
   \subfigure[$\SIR$ coverage]{\centering 
     \includegraphics[width=8cm]{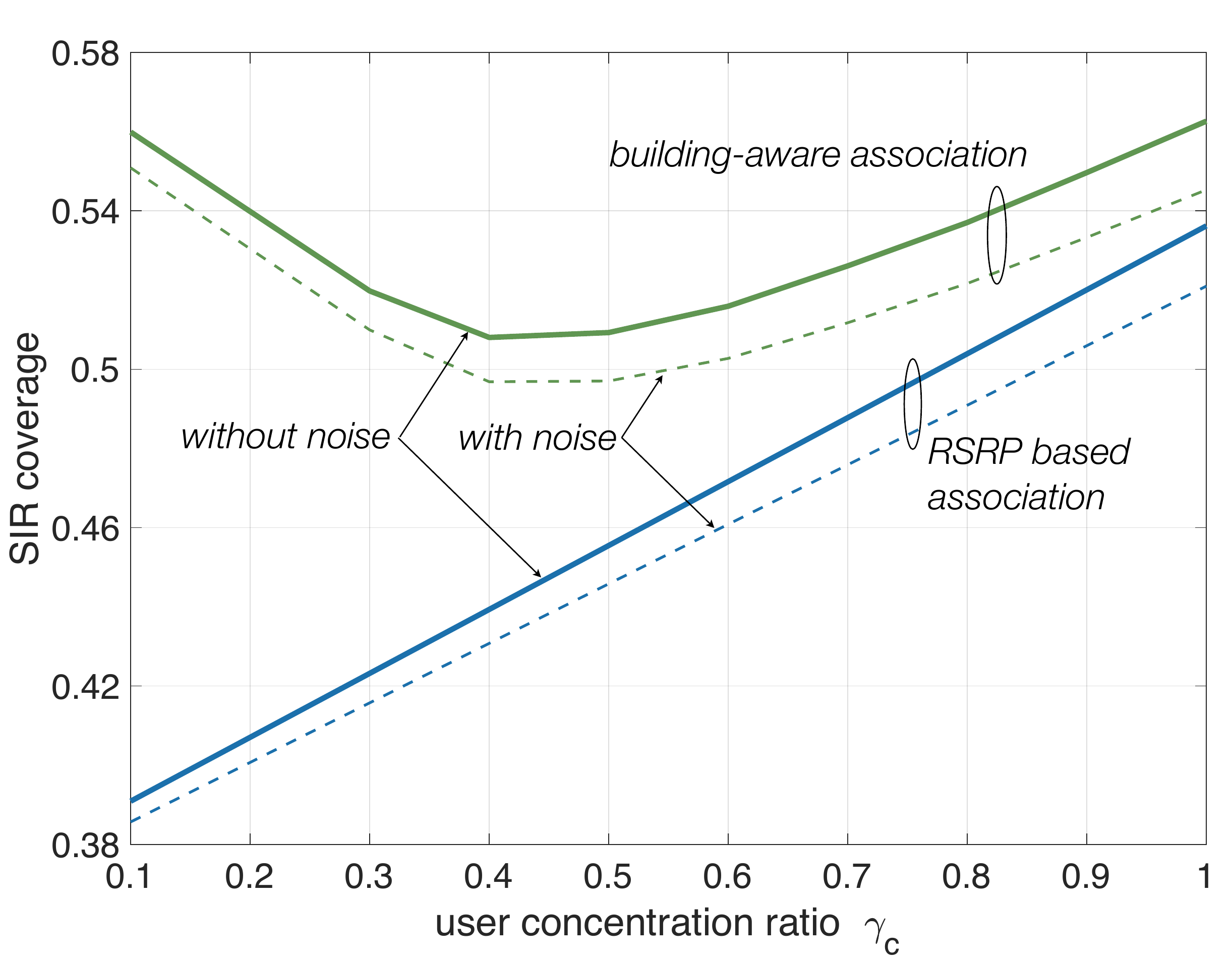} }
   \subfigure[average rate]{
     \includegraphics[width=8cm]{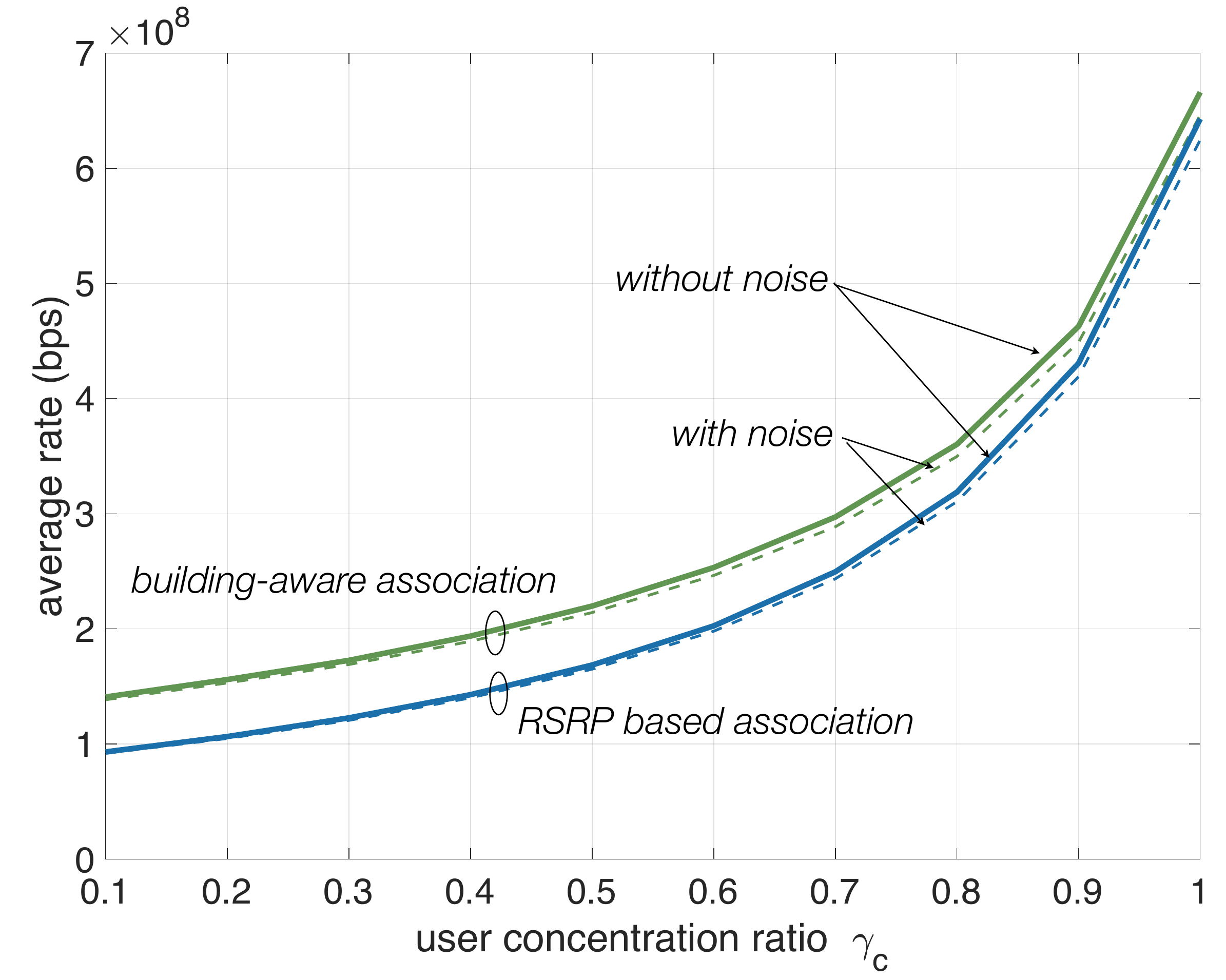}}
\caption{Impact of user concentration ratio $\gamma_c$ on 
$\SIR$ coverage and rate ($\lambda_b= \text{600 BSs}/\text{km}^2$, $\lambda= \text{2 $\times 10^3$ UEs}/\text{km}^2$, $\lambda_\ell= \text{$500$ buildings}/\text{km}^2$, $\theta=\frac{\pi}{4}$).}\label{Fig:conc}
\end{figure*}

This Corollary provides the building-aware association design guideline for a special scenario that is asymptotic but highly in accordance with the BS densification trend, where the number of BSs exceed the UE number and some BSs may not serve any UE \cite{JH15,JH_globecom}.
By regarding ${\beta_{\mathcal{R}}}^*={\beta_{\mathcal{S}}}^*$, we can efficiently diminish the calculation time due to the smaller searching range needed to find the optimal $\beta$.
The accuracy of Corollary \ref{asym_prop} is verified in Fig.~\ref{Fig:bias}.

In addition to this asymptotic case, the mathematical form in (\ref{optbeta_r}) provides a useful design guideline.
This stems from the intuition that there is an optimal ratio of D-BSs among the BSs in the LOS region to maximize the rate by jointly optimizing the $\SIR$ and the amount of available resources.
Keeping this optimal ratio in mind, we summarize the building-aware association design guideline according to different network parameters, as in the following remarks.

\begin{remark}{The optimal bias ${\beta_\mathcal{R}}^*$ is affected by the network parameters as follows:
\begin{enumerate}[$\hspace{10pt}1.$]
\item{As the building density and/or length decreases, the optimal bias ${\beta_{\mathcal{R}}}^*$ should be increased.}
\item{As the beamwidth $\theta$ decreases, the optimal bias ${\beta_{\mathcal{R}}}^*$ should be decreased.
}
\end{enumerate}
} 
\end{remark}

The reason behind is that as the building density and/or length decreases, the LOS region increases, resulting in an increment in the number of LOS BSs. 
Therefore, the optimal bias ${\beta_{\mathcal{R}}}^*$ should be increased to keep the optimal ratio of D-BSs among the BSs in the LOS region.
On the otherhand, ${\beta_{\mathcal{R}}}^*$ increases with the beamwidth.
This is because, when BSs can transmit sharper beam signals, less BS beam signals leak beyond their nearest building, i.e. more BSs become D-BS by satisfying the D-BS condition \eqref{condition} in Section II-D. Therefore, the optimal bias ${\beta_{\mathcal{R}}}^*$ should be decreased to keep the optimal ratio of D-BSs among BSs in the LOS region.


\section{Numerical Evaluations of Building-aware Association Algorithm}
\label{sec:simul}

The $\SIR$ coverage and rate of building-aware association scheme (Section III) and the corresponding optimal bias (Section IV) are evaluated in this section.
In addition, the practical viability is validated by applying a real-world building geography in three cities, Gangnam, Manhattan, and Chicago.

\subsection{$\SIR$ Coverage and Rate according to Different System Parameters}

Figs.~\ref{Fig:build_den}-\ref{Fig:conc} visualize the $\SIR$ coverage and rate, according to the building density, beamwidth, and user concentration ratio, respectively.
Default simulation parameters are given as follows: $W=500$ MHz \cite{bwref}, $d_l=30$ m, $d_l=10$ m, $g_m=20$ dB, $g_s=0$ dB, $t=10$ dB, transmit power $=23$ dBm, and noise power $=-77$ dBm according to the following equation: $-174 \text{ dBm/Hz} + 10 \log_{10}(W \text{ Hz}) + 10 \text{ dB}$.
The association bias $\beta$ by default is set to be the optimal value.

\begin{table}[t]\caption{Building Statistics \cite{JH15}} \centering
	\begin{tabular}{c c c c  }
	\hline Building parameters (unit)  & Manhattan& Gangnam & Chicago \\
	\thickhline 
	Density (buildings/${\text{km}}^2$)  &  $1467$ & $1010$ &$474$\\
	Average length (m)  & $26.5$ & $22.41$& $36.35$ \\ 
	Average width (m)  & $20.83 $ & $9.35$ & $21.48$\\
	Average LOS distance (m)  & $23.12$& $62.40$ & $69.74$
	\\ 
	 \hline
\end{tabular}
\label{table:buildings}
\end{table}
\begin{figure*}     
\centering
   \subfigure[$\SIR$ coverage gain]{\centering 
     \includegraphics[width=8cm]{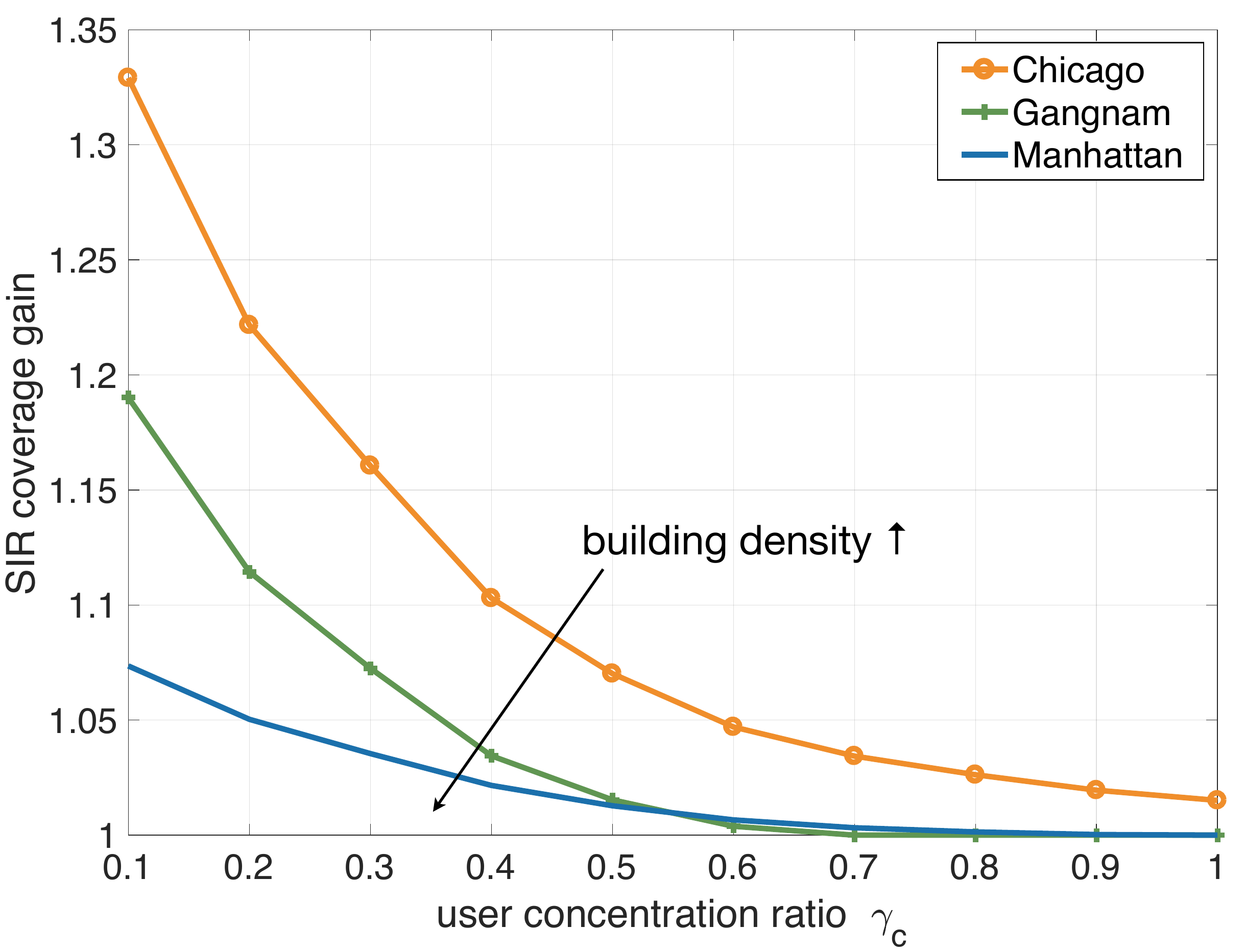} }
   \subfigure[rate gain]{
     \includegraphics[width=8cm]{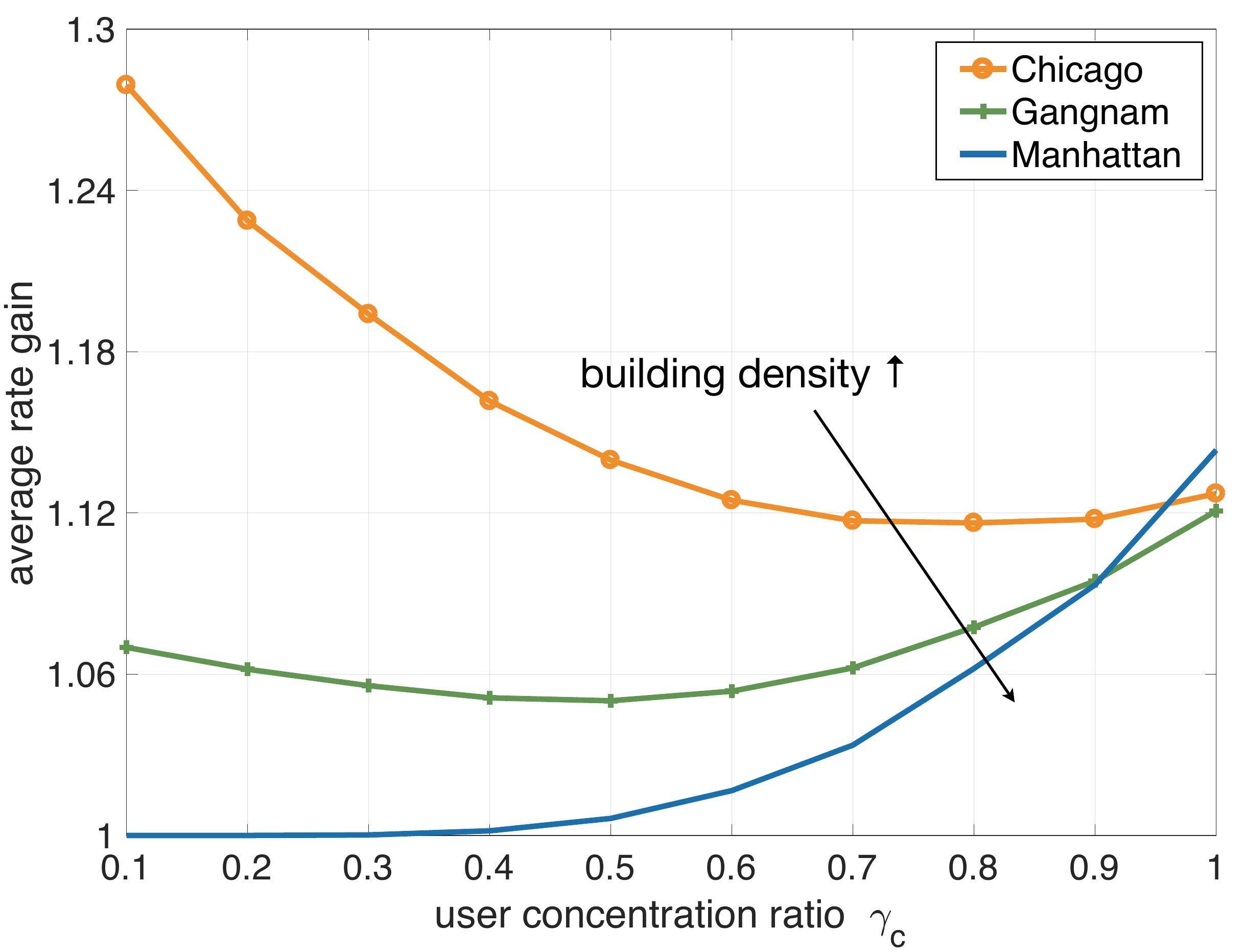}}
\caption{ $\SIR$ coverage and rate gain according to the user concentration ratio $\gamma_c$ ($\lambda_b= \text{500 BSs}/\text{km}^2$, $\lambda= \text{20 $\times 10^3$ UEs}/\text{km}^2$, $\theta=\frac{\pi}{6} rad$).} \label{Fig:geo_gain}
\end{figure*}

\begin{figure*}     
\centering
   \subfigure[$\SIR$ coverage optimal bias]{\centering 
     \includegraphics[width=8cm]{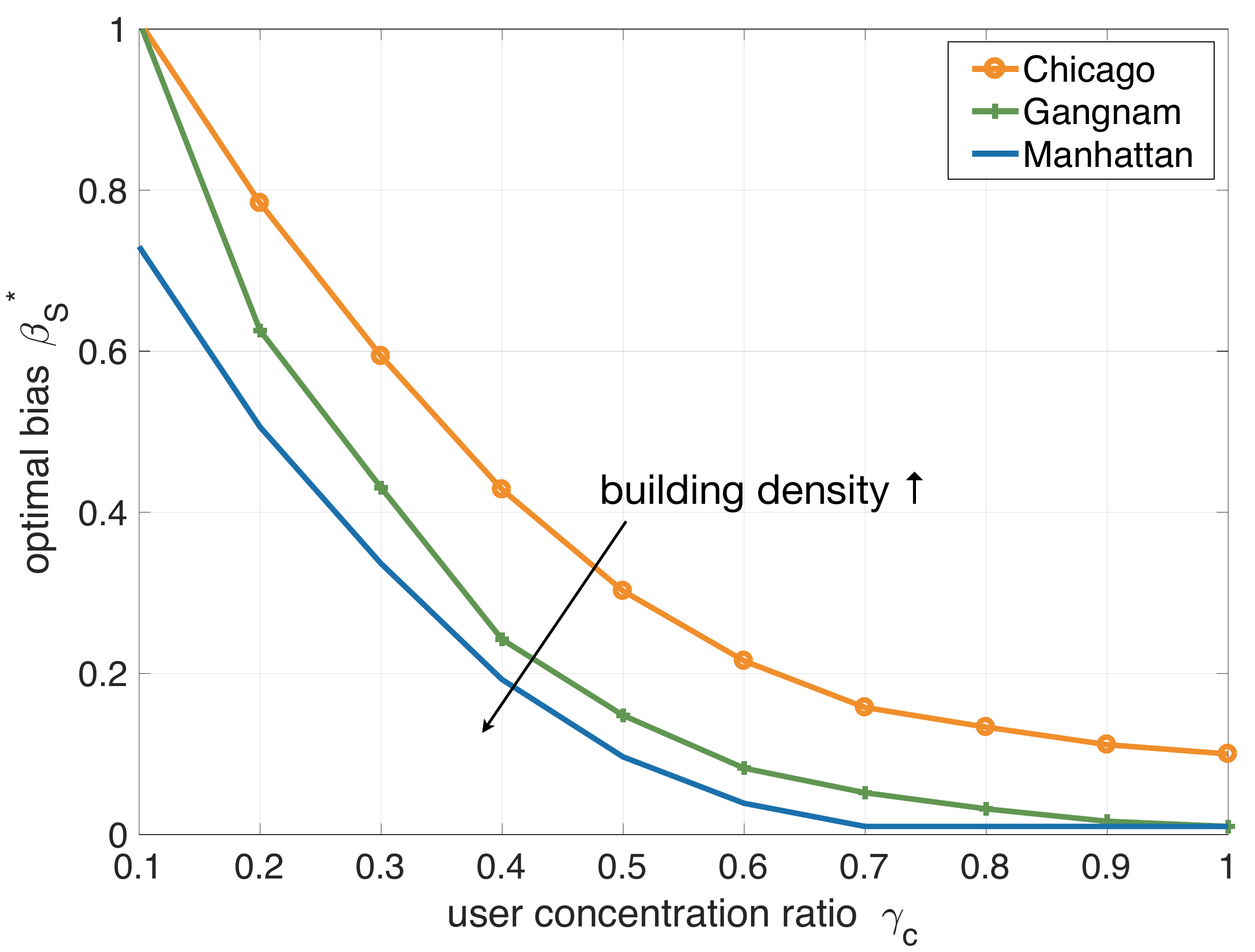} }
   \subfigure[rate optimal bias]{
     \includegraphics[width=8cm]{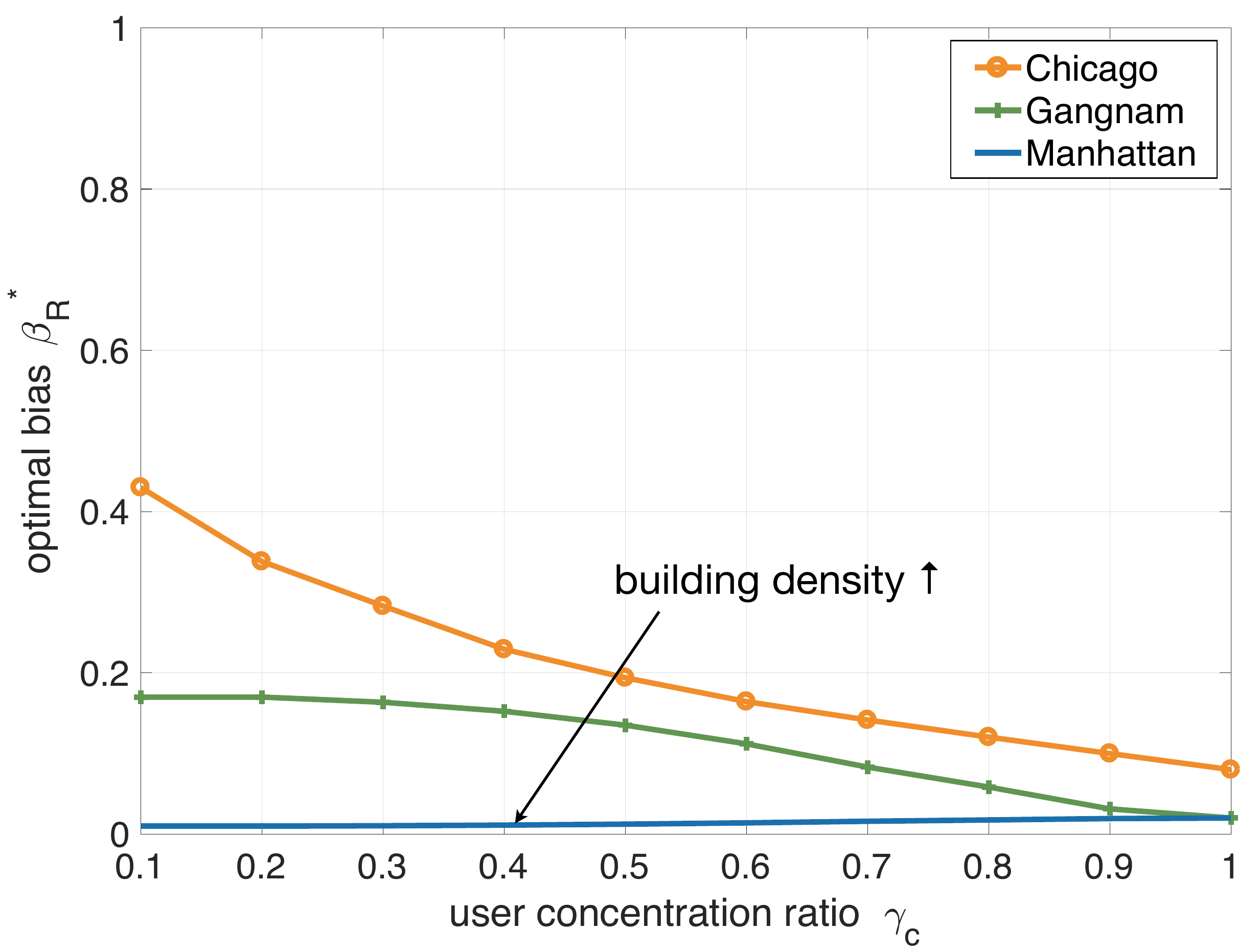}}
\caption{Optimal biases for maximizing $\SIR$ coverage and rate according to the user concentration ratio $\gamma_c$ ($\lambda_b= \text{500 BSs}/\text{km}^2$, $\lambda= \text{20 $\times 10^3$ UEs}/\text{km}^2$, $\theta=\frac{\pi}{6} rad$).} \label{Fig:optbias}
\end{figure*}

The three results show that the building-aware association scheme can achieve a superior $\SIR$ coverage and rate compared to the RSRP-based association scheme.
In addition, they also illustrate the effect of considering noise power (see the dotted lines).
To this end, we additionally calculate the $\mathsf{SINR}$ coverage with minor modifications from the $\SIR$ coverage \eqref{Eq:SIRcov}, by deviding the $\mathsf{SINR}$ coverage into $\mathsf{SNR}$ and $\SIR$ terms independently \cite{tractable}.
When considering the noise power, both the $\SIR$ coverage and rate decrease, but they still follow a similar trend as the results that do not consider the noise power, thus justifying the noise power elimination in our analysis.

Fig.~\ref{Fig:build_den} indicates that the gain of our algorithm does not monotonically increase with the building density $\lambda_\ell$.
This relates to the number of BSs that become D-BS using the algorithm.
When $\lambda_\ell$ is small, there is not a sufficient number of BSs around the buildings, reducing the number of BSs that can become D-BS.
The gain thus increases with $\lambda_\ell$ at first.
However, when $\lambda_\ell$ is high, the number of BSs in the LOS region decreases due to the shrinkage of the LOS region.
This also reduces the number of D-BS candidates, saturating the rate improvement.
Fig.~\ref{Fig:beam} shows that the building-aware association algorithm can complement the limitations of beam-forming technology. 
For instance, the building-aware association rate when $\theta = 0.3 \pi$ is similar to that of the RSRP based association scheme when $\theta = 0.15 \pi$, implying that the number of antennas can be reduced.
Fig.~\ref{Fig:conc} demonstrates that the $\SIR$ coverage gain using the building-aware association scheme decreases with the user concentration ratio $\gamma_c$ whereas the rate gain does not.
This is because the amount of $\SIR$ increment due to interference avoidance is high when most of the UEs are far from buildings (see Remark 1).
On the other hand, the amount of rate increment by balancing the traffic load is high when most of UEs are adjacent to buildings so that the traffic unbalance is severe.
Such load balancing gain compensates for the $\SIR$ gain decrement when $\gamma_c$ is high.




\subsection{Building-aware Association under a Real Building Geography}

Using previous studies \cite{JH15} and the open source geographic information, we calculate the building parameters for Gangnam, Manhattan, and Chicago as summarized in Table II.

Fig.~\ref{Fig:geo_gain} shows that the gain using the building-aware association scheme in Chicago is higher than that in Gangnam or Manhattan, implying that our algorithm is profitable in network geography like Chicago.
The main reason behind is the building density $\lambda_\ell$ is smaller in Chicago compared to other cities.
The corresponding large LOS region enables a sufficient number of BSs to become D-BS, leading to a further improvement in the average rate.
As shown in Fig.~\ref{Fig:build_den} above, when $\lambda_\ell$ is too small or large the gain of proposed algorithm decreases.
Besides, it is worth noting that the average building length in Chicago is longer than that in Gangnam and Manhattan.
This implies that when the building length is long, there are more D-BS candidates around the buildings, increasing the feasible set of $\beta$ to be optimized.

Although the rate gain in Manhattan is low compared to the gain in other cities, the average rate gain increases with the user concentration ratio $\gamma_c$.
When UE is more frequently concentrated around buildings such as in dense urban scenarios like Manhattan, the result implies that our algorithm can still guarantee a large improvement.

Fig.~\ref{Fig:optbias} illustrates that the optimal bias for the Manhattan scenario is low compared to that of the Gangnam and Chicago scenarios.
This is mainly due to the fact that the LOS distance in Manhattan is short, thus there are a smaller number of BSs in the LOS region.
Therefore, the optimal bias should be decreased to keep the optimal ratio of the number of D-BSs among BSs in the LOS region, as explained in Remark 3.
This figure also shows that the optimal bias ${\beta_\mathcal{S}}^*$ is $1$ when the ratio $\gamma_c=0$ and then the optimal bias decreases.

%
%
%


\section{Conclusion}

This study addresses the mmW interference problem due to the mmW BS densification and amplified directional signal strength.
As a solution, we propose a building-aware association scheme where BSs adjacent to buildings only associate with UEs toward a building.
The impact of this scheme is analyzed in terms of the average data rate using stochastic geometry (Proposition $2$), verifying large improvements compared with the RSRP-based association scheme.
The result indicates that the rate improvement is convex-shaped over the building density.
In addition, it sheds light on obtaining the optimal association bias $\beta$ that jointly optimizes the interference decrease and load balancing (Proposition $3$).
The result implies that the optimal bias $\beta^*$ should be decreased with the building density and length, while it should be increased with the beamwidth.
Real geography based blockage models in Gangnam, Manhattan and Chicago are used to validate the practical feasibility of the proposed algorithm, showing that our algorithm provides a higher rate improvement in Chicago scenario which has less buildings compared to other cities.

Further extension could contemplate the impact of UE mobility on traffic usage patterns.
The traffic amount is known to have a convex-shape over the UE velocity \cite{convexity}.
Considering that UEs are likely to move at higher speeds when they are away from buildings (e.g., in-vehicle passengers), UEs far from buildings might cause more data traffic.
By taking into account this behavior, we can provide an optimal association bias that better suits practical networks.

\section{Appendix}

\subsection{Proof of Lemma 1}
In the average LOS ball model \cite{Bai14}, a typical UE considers that there is a rectangular blockage at a distance $R_L$ from it in all directions, implying that there are theoretically an infinite number of rectangular blockages at a distance $R_L$.
Recall that a BS becomes a D-BS when its beam signal does not exceed the nearest building with length $\beta\L$.
Since the circumferential angles in the same arc of a circle are all equal, the region of D-BSs becomes a circle truncated by the building, where the circumferential angle is $\theta$ (see Fig.~\ref{bl_region}).
Then the distance from the building to its farthest D-BS becomes $\frac{\beta\L}{2 \tan\(\frac{\theta}{2}\)}$ using trigonometric properties.
Since each blockage turns each BS within the maximum distance $\frac{\beta d_l}{2\tan{\frac{\theta}{2}}}$ from it to a D-BS and there are infinite blockages, the modified LOS area shrinks to have a radius of $R_\beta= \max \(R_L - \frac{\beta d_l}{2\tan{\frac{\theta}{2}}},0\)$.

\subsection{{Proof of Proposition 1}}

\begin{figure}     
\centering
   \subfigure[Region of D-BSs]{\centering
     \includegraphics[width=4cm]{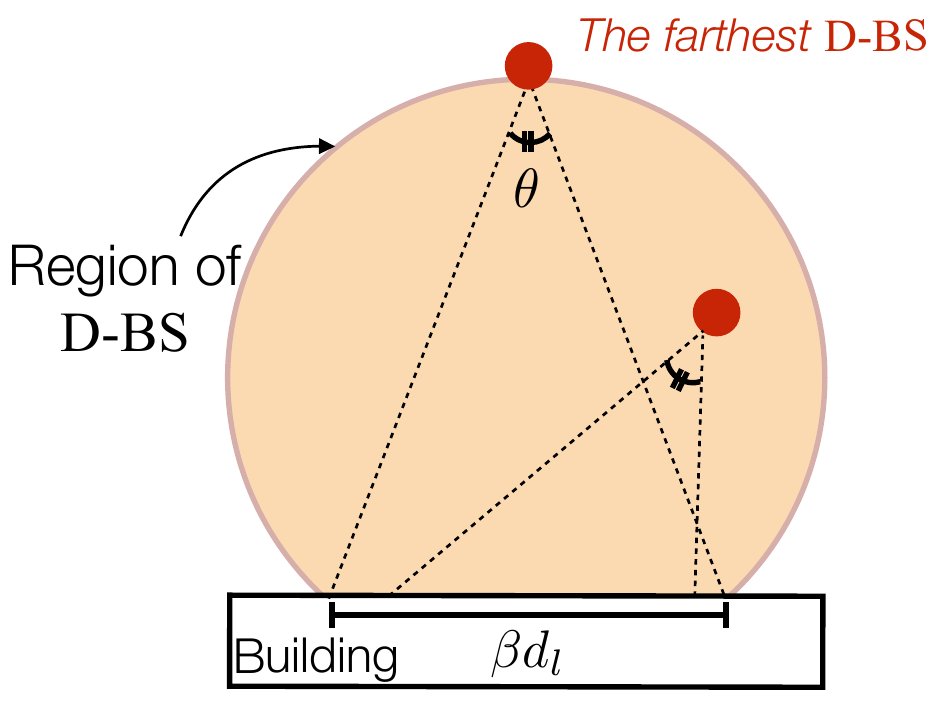}  \label{bl_region}}
   \subfigure[$U_n$ LOS region division]{
     \includegraphics[width=4cm]{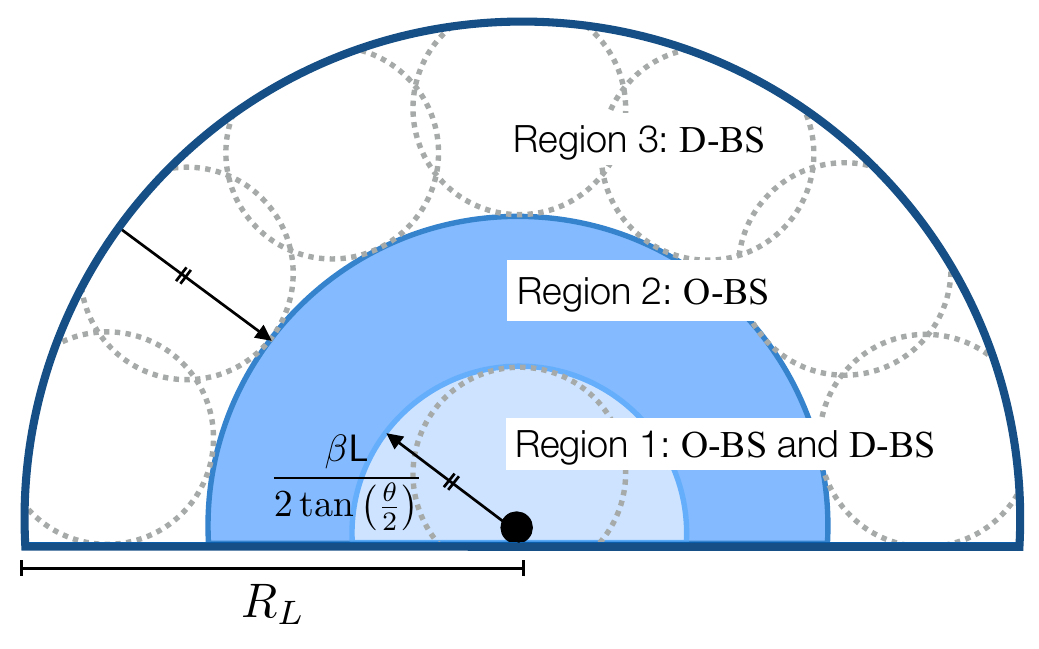} \label{U_n_intf}} 
\caption{Illustrations of D-BS condition (a) and region of D-BSs and O-BSs from the perspective of a typical $U_n$ (b).} 
\end{figure}

To derive the average $\SIR$ coverage, we calculate the $\SIR$ coverage of a typical $U_r$ and that of a typical $U_n$ respectively.
At first, $\mathcal{S}_r(\beta)$, the $\SIR$ coverage of $U_r$ can be achieved through the following preliminary techniques.
\begin{enumerate}
\item{\textbf{Distance Distribution under LOS Condition.}
Define $R$ as $R_0$ conditioned on $R_0 \leq R_L$. The cumulative density function (CDF) of $R$ is 
\begin{align}\nn
\mathsf{P}\(R>r\) &:= \mathsf{P}\(R_0>r | R_0 \leq R_L\) 
= \frac{\mathsf{P}\(r<R_0\leq R_L\)}{\mathsf{P}\(R_0 \leq R_L\)}\\
 &= \frac{\mathsf{P}\(R_0>r\)-\mathsf{P}\(R_0>R_L\)}{\mathsf{P}\(R_0 \leq R_L\)}.\label{CDF_cond}
\end{align}\normalsize
By differentiating (\ref{CDF_cond}), we can get the PDF of $R$: $f_R(r) = \frac{f_{R_0}(r)}{\mathsf{P}\(R_0 \leq R_L\)},$ 
where $f_{R_0}(r)=2\pi\lambda_b r \exp(-\pi\lambda_b r^2).$
}
\item{\textbf{Directional Interference Thinning.}
Now that BSs transmit directional signal, the interfering BSs are decomposed into two groups: who transmit interfering signals with the antenna gain $g_m$ and $g_s$.
Since the probability that angle from a BS is within the beamwidth $\theta$ is $\frac{\theta}{2\pi}$, each BS's link to a typical user has a gain $g_m$ with probability $\frac{\theta}{2\pi}$ and $g_s$ with probability $1-\frac{\theta}{2\pi}$.
According to the mapping theorem, the power decrease with $\frac{g_s}{g_m}$ has an equal effect with density decrease with $\(\frac{g_s}{g_m}\)^{\frac{2}{\alpha}}$ \cite{JH15}.
The actual interferer thus can be thinned with a probability $p_a=\frac{\theta}{2\pi} +\frac{2\pi-\theta}{2\pi} {\(\frac{g_s}{g_m}\)}^{\frac{2}{\alpha}}$.
}
\end{enumerate}

The $\SIR$ coverage then is represented as below.
\begin{small}\begin{align}\nn
&\mathcal{S}_r(\beta) \\
&=\mathsf{P}\(R_0 \leq R_L\) \int_0^{R_L} \mathsf{P}\(\frac{g_m r^{-\alpha} h }{\sum_{x_i\in \Phi_{{b}}\(\lambda_b \)}{G_i {|x_i|}^{-\alpha} h_i}}> t \) f_R(r) dr \\ \nn
&\overset{(a)}{=} \int_0^{R_\beta} \mathsf{P}\(\frac{g_m r^{-\alpha} h }{\sum_{x_i\in \Phi_{{b}}\(\lambda_b \)}{G_i {|x_i|}^{-\alpha} h_i}}> t \) f_{R_0}(r) dr \\
&\quad+ \int_{R_\beta}^{R_L} \mathsf{P}\(\frac{r^{-\alpha} h }{\sum_{x_i\in \Phi_{{b}}\(\lambda_b \)}{ \frac{g_s}{g_m}{|x_i|}^{-\alpha} h_i}}> t \) f_{R_0}(r) dr \\
\nn &\overset{(b)}{=}\int_0^{R_\beta} \exp\(-{\pi\lambda_b r^2} \[p_a t^{\frac{2}{\alpha}} \int_{t^{-\frac{2}{\alpha}}}^{\frac{{R_\beta}^2}{r^2 t^{\frac{2}{\alpha}}}} {\frac{du}{1+u^{\frac{\alpha}{2}}}} \right.\right.\\
\nn &\quad \left. \left. + \(\frac{g_s t}{g_m}\)^{\frac{2}{\alpha}} \int_{\frac{{R_\beta}^2}{r^2 t^{\frac{2}{\alpha}}}}^{\frac{{R_L}^2}{r^2 t^{\frac{2}{\alpha}}}} {\frac{du}{1+u^{\frac{\alpha}{2}}}}\]\) f_{R_0}(r) dr \\
&\quad +\int_{R_\beta}^{R_L} {\exp\(-{\pi\lambda_b r^2} \[\(\frac{g_s t}{g_m}\)^{\frac{2}{\alpha}} \int_{t^{-\frac{2}{\alpha}}}^{\frac{{R_L}^2}{r^2 t^{\frac{2}{\alpha}}}} {\frac{du}{1+u^{\frac{\alpha}{2}}}}\]\) f_{R_0}(r)} dr,\label{s_n_appen}
\end{align}
\end{small}
where \small$R_1=\min\(R_L-R_\beta,\frac{R_L}{2}\)$, $r_1 = {\max\[\max\(R_\beta,\frac{R_L}{2}\)^2,r^2\]}{\(r t^{\frac{1}{\alpha}}\)^{-2}}$, $r_2 = {\min\(R_L-R_\beta,\frac{R_L}{2}\)^2}{\(r t^{\frac{1}{\alpha}}\)^{-2}}$\normalsize.
Step $(a)$ follows from that when the signal distance is longer than $R_\beta$, a typical $U_r$ receives interference signal only with a side lobe gain $g_s$.
In addition, step $(b)$ follows from: (i) directional interference thinning with mapping theorem and (ii) the modified LOS region.
By applying $f_{R_0}(r)$ and integrating it from $R_\beta$ to $R_L$, we can derive the $\mathcal{S}_r(\beta) $.

The $\SIR$ coverage of a typical $U_n$ differs from that of a typical $U_r$ because their LOS regions are different and some D-BSs can interfere a $U_n$.
\begin{enumerate}
\item{\textbf{Distance Distribution.} Remind that a typical $U_n$ has a half moon shaped LOS region due to the building on its side. 
This implies that a $U_n$ cannot receive signals from BSs on the other side of the building, making the signal distance PDF different as below.
\begin{align}
f_{R_0}(r) = \pi\lambda_b r \exp \(-\frac{\pi\lambda_b r^2}{2}\). \label{distpdf_nc}
\end{align}}
\item{\textbf{Interference Division.} In order to calculate the interference of a typical $U_n$, we separate the LOS region into three regions as shown in Fig.~\ref{U_n_intf}. Firstly, in region 1, there are D-BSs and O-BSs. Since D-BSs in this region transmit signals toward the building where the $U_n$ is attached, we assume that their signals always interfere with $U_n$. Secondly, in region 2, there are only O-BSs. They interfere with $U_n$ if their signal directions are toward the $U_n$. Thirdly, in region 3, there are only D-BSs who transmit signals toward UEs on their nearest buildings side, causing side lobe interference to the $U_n$.
} 
\item{\textbf{Probability of Being Interferer in Region 1.} Since the probability of a BS in region 1 interfering with the $U_n$ depends on whether it is a D-BS or a O-BS, it is important to derive the probability of a BS being D-BS. 
The probability is derived by the ratio of the area of region 1 to the truncated circle inside region 1 as follows: \small$\(\frac{(\pi-\theta)^2}{4 \sin^2(\theta)}+\frac{1}{4 \tan (\theta)}\)\frac{8 \tan^2\(\frac{\theta}{2}\)}{\pi}.$\normalsize
By considering this probability and main lobe interfering probability, the probability $P_\ell$ that a BS in region 1 becomes a main lobe interferer is
\small$P_\ell=\(\frac{(\pi-\theta)^2}{4 \sin^2(\theta)}+\frac{1}{4 \tan (\theta)}\)\frac{8 \tan^2\(\frac{\theta}{2}\)}{\pi}+\[1-\(\frac{(\pi-\theta)^2}{4 \sin^2(\theta)}+\frac{1}{4 \tan (\theta)}\)\frac{8 \tan^2\(\frac{\theta}{2}\)}{\pi} \] p_a.$\normalsize}
\end{enumerate}
By utilizing the distance distribution and interference division, we can derive the $\SIR$ coverage of a typical $U_n$ as follows.
\begin{small}{\begin{align} \nn
&\mathcal{S}_n(\beta) \\
&=\mathsf{P}\(R_0 \leq R_L\) \int_0^{R_L} \mathsf{P}\(\frac{g_m r^{-\alpha} h }{\sum_{x_i\in \Phi_{{b}}\(\lambda_b\)}{G_i {|x_i|}^{-\alpha} h_i}}> t \) f_R(r) dr \\ \nn
&=  \int_{\max\(R_\beta,\frac{R_L}{2}\)}^{R_L}  \mathsf{P}\(\frac{r^{-\alpha} h }{\sum_{x_i\in \Phi_{{b}}\(\lambda_b\)}{\frac{g_s}{g_m}{|x_i|}^{-\alpha} h_i}}> t \) f_{R_0}(r) dr \\
&+  \int_0^{\max\(R_\beta,\frac{R_L}{2}\)} \mathsf{P}\(\frac{g_m r^{-\alpha} h }{\sum_{x_i\in \Phi_{{b}}\(\lambda_b\)}{G_i{|x_i|}^{-\alpha} h_i}}> t \) f_{R_0}(r) dr \\ 
\nn &\overset{(a)}{=} \int_{R_1}^{R_L} \pi\lambda_b r \exp \[-\frac{\pi\lambda_b r^2}{2} \(1+p_a t^{\frac{2}{\alpha}} \int_{t^{-\frac{2}{\alpha}}}^{{r_1}} \frac{du}{1+u^{\frac{\alpha}{2}}} \right. \right. \\ \nn
&\quad \left.\left.+ \[\frac{g_s t}{g_m}\]^{\frac{2}{\alpha}} \int_{{r_1}}^{\frac{{R_L}^2}{r^2 t^{\frac{2}{\alpha}}}}\frac{du}{1+u^{\frac{\alpha}{2}}}\)\] dr  \\ \nn
&\quad+ \int_0^{R_1} \pi\lambda_b r \exp\[-\frac{\pi\lambda_b r^2}{2} \(1+P_\ell t^{\frac{2}{\alpha}} \int_{t^{-\frac{2}{\alpha}}}^{{r_2}} \frac{du}{1+u^{\frac{\alpha}{2}}}\right.\right. \\
&\quad \left.\left.+ p_a t^{\frac{2}{\alpha}} \int_{{r_2}}^{{r_1 }}\frac{du}{1+u^{\frac{\alpha}{2}}}+ \[\frac{g_s t}{g_m}\]^{\frac{2}{\alpha}} \int_{{r_1}}^{\frac{{R_L}^2}{r^2 t^{\frac{2}{\alpha}}}}\frac{du}{1+u^{\frac{\alpha}{2}}}\)\]  dr, \label{s_c_appen}
\end{align} }\end{small}
where $(a)$ follows from the division of interference region. When the association distance $r$ is shorter than $\min\(R_L-R_\beta, \frac{R_L}{2}\)$, D-BS's in region 1 and O-BSs in region 1 and 2 can interfere with the $U_n$ but their corresponding probabilities of being interferer are different.

Note that the path-loss exponent $\alpha$ in mmW networks is small compared to that used in the conventional sub-$6$ GHz frequency since we only consider LOS communication links \cite{pathloss}.
Considering that $\alpha$ cannot be smaller than $2$ when we observe a $2$-dimensional space \cite{kendall}, we only focus on the case that $\alpha$ larger than $2$ goes to it, i.e. $\alpha \to 2+$.
For the exponent $\alpha$ close to $2$, we can remove double integration in \eqref{s_n_appen} and \eqref{s_c_appen} by using a property $\lim_{\alpha \to 2}\int {\frac{dx}{1+x^{\frac{\alpha}{2}}}} = \ln \(1+x\)$.
Then the $\SIR$ coverage $\mathcal{S}_r(\beta)  $ \eqref{s_n_appen} is simplified as below.
{\begin{small}\medmuskip=2mu\thinmuskip=2mu\thickmuskip=2mu 
\begin{align} \nonumber
&\lim_{\alpha \to {2}+} \mathcal{S}_r(\beta)  = 
2F_0^{R_\beta} \(2+2 p_a t\int_{t^{-1}}^{\frac{{R_\beta}^2}{r^2 t}} {\frac{du}{1+u}}+ \frac{2 g_s t}{g_m}\int_{\frac{{R_\beta}^2}{r^2 t}}^{\frac{{R_L}^2}{r^2 t}} {\frac{du}{1+u}}\)  \\&\quad+ 2F_{R_\beta}^{R_L} \(2+ \frac{2 g_s t}{g_m} \int_{t^{-1}}^{\frac{{R_L}^2}{r^2 t}} {\frac{du}{1+u}}\) \\ \nn
&\quad =2F_0^{R_\beta}\[2+{2 p_a t }\ln \(\frac{t+{R_\beta}^2 r^{-2}}{1+t}\)+{ \frac{2 g_s t}{g_m}  }\ln \(\frac{t+{R_L}^2 r^{-2}}{t+{R_\beta}^2 r^{-2}}\)\] \\
&\quad +2F_{R_\beta}^{R_L} \[2+{ \frac{2 g_s t}{g_m}  } \ln \(\frac{t+{R_L}^2 r^{-2}}{t+1}\)\]
 \end{align} \end{small}}
 The same process is applied to derive $\mathcal{S}_n(\beta)$ .

By applying $\mathcal{S}_r(\beta)$ and $\mathcal{S}_n(\beta)$ into $\mathsf{P} \(\SIR>t\) = \(1-\gamma_c\) \mathcal{S}_r(\beta)  + \gamma_c \mathcal{S}_n(\beta) $, we can finalize the proof. \hfill $\blacksquare$

\subsection{Proof of Proposition 2}
We need to calculate the average UE number in the cell coverage for a typical UE $N_n$ and $N_r$.
\subsubsection{Derivation of $N_r$}
When $\beta=0$, i.e. there are no D-BS, $N_r=\frac{1.28\lambda_r}{\lambda}$ since a typical cell is 1.28 times larger than other cells on average \cite{offloading} and the typical cell is located near the center of the LOS ball and assumed to have no $U_n$'s who are attached to buildings.
As $\beta$ increases, O-BSs should expand their association areas when their neighboring BSs become D-BS and shrink their coverage areas.
To check if any neighboring BSs of a typical cell become D-BS, we consider that the average number of neighboring BSs of a typical BS in two-dimensions is $6$ \cite{avg_edge} and the average distance to the $6$-th nearest neighboring node from a typical $U_r$ is $\frac{0.68}{\sqrt{\lambda}}$ \cite{kendall}.
For computational brevity, we assume that the typical cell expands its cell area when $\frac{0.68}{\sqrt{\lambda}}>R_\beta$, i.e. when its $6$-th nearest neighboring BS becomes D-BS.
Otherwise, it keeps its association area.

To derive the average expanded area of the typical cell, we need to calculate the area of an O-BS.
The area of an O-BS is derived by assuming that the association region of a D-BS is equal to a triangle whose base line length is $\beta \L$ and height is the distance to the building.
\begin{small}\begin{align} \nn
 \mathsf{E} &\ \(\pi {R_L}^2 - \sum_{R_\beta \leq r_i \leq R_L} \frac{\beta \L}{2} \(R_L-r_i\)\) \\
 &= \pi {R_L}^2 - \frac{\beta \L}{2} \int_{R_\beta}^{R_L} 2\pi\lambda_b r (R_L-r) dr \\
&= \pi {R_L}^2 - \frac{\beta \L}{2} \[\pi\lambda_b R_L \({R_L}^2 - {R_\beta}^2\) - \frac{2 \pi\lambda_b \({R_L}^3-{R_\beta}^3\)}{3}\].
\end{align}
\end{small}
\noindent Since we do not consider the overlapping region among D-BS association regions, the above area may be smaller than $\pi{R_\beta}^2$.
Thus we derive the average area of O-BS as below. 
\begin{small}\begin{align} \nn
A_c = \max &\(\pi{R_\beta}^2, \pi {R_L}^2 - \frac{\beta \L}{2} \[\pi\lambda_b R_L \({R_L}^2- {R_\beta}^2\) \right.\right. \\
&\left.\left. - \frac{2}{3}\pi\lambda_b \({R_L}^3-{R_\beta}^3\) \] \).
\end{align}\end{small}

\noindent Since the association area of an O-BS is expanded from $\pi {R_\beta}^2$ to $A_c$, the average association area of the typical cell becomes $\frac{1.28 A_c}{\lambda_b \pi {R_\beta}^2}$.
Within the area of an O-BS, there are $U_r$'s and $U_n$'s.
To derive the average UE number in the typical cell coverage, we calculate the area of $U_r$'s $A_r$.
If $d_c > R_L-R_\beta$, the area $A_r$ becomes $\pi \(R_L - d_c\)^2$.
Otherwise, by calculating $\mathsf{E} \[\sum_{R_\beta \leq r_i \leq R_L-d_c} \frac{\beta \L}{2} \(R_L-r_i-d_c\)\]$, the area $A_r$ becomes 
\begin{small}\begin{align}\nn
A_r=\max&\(\pi{R_\beta}^2,\pi \[R_L-d_c\]^2-\frac{\beta\L\pi\lambda}{2}\[ \(R_L-d_c\)\(\[R_L-d_c\]^2 \right.\right.\right.\\ 
&\left.\left.\left.-{R_\beta}^2\)-\frac{2\(\[R_L-d_c\]^3-\[R_\beta-d_c\]^3\)}{3}\]\).
\end{align}\end{small} 
Then the average UE number in the typical cell $N_r$ becomes $\frac{1.28\[\(A_c-A_r\)\lambda_n + A_r \lambda_r\]}{\lambda_b \pi{R_\beta}^2}$.

\subsubsection{Derivation of $N_n$}
The area of a typical cell depends on the association distance $r$.
When $0 \leq r < d_c$, there are only $U_n$'s in the association area of the typical cell. Since the associated BS is D-BS and its average cell area is $\frac{1.28\beta\L r}{2}$, the average UE number in the typical cell becomes $\frac{1.28\beta\L r \lambda_n}{2}$.
When $d_c \leq r < \min\(R_L-R_\beta,\frac{R_L}{2}\)$, the associated BS is still a D-BS and its average cell area is $\frac{1.28\beta\L r}{2}$ but there are both of $U_n$'s and $U_r$'s. So the average UE numbers becomes ${1.28\beta \L} \(\frac{r-d_c}{2} \lambda_r + d_c \lambda_n\)$.
When $ \min\(R_L-R_\beta,\frac{R_L}{2}\) \leq r \leq R_L$, the BS could be a D-BS or not, and the average UE number thus is similar to $N_r$.
By applying the signal distance distribution of $N_n$ \eqref{distpdf_nc}, we can finalize the proof.  \hfill $\blacksquare$

\subsection{Proof of Proposition 3}
If $\beta > \frac{R_L \tan \(\frac{\theta}{2}\)}{\L} $, the coverage $\mathcal{S}_n(\beta) $ no longer depends on $\beta$. 
Also we know that the coverage $\mathcal{S}_r(\beta) $ is monotonically increasing with $\beta$.
The optimal $\beta$ in the interval from $\frac{R_L \tan \(\frac{\theta}{2}\)}{\L}$ to $1$ thus become $1$.
What we have to do is then to compare $ \gamma_c \mathcal{S}_n\(\frac{R_L \tan \(\frac{\theta}{2}\)}{\L}\) + (1-\gamma_c) \mathcal{S}_r(1)$ and $\max_{0 \leq \beta \leq {R_L \tan \(\frac{\theta}{2}\)}{\L}^{-1}} \gamma_c \mathcal{S}_n(\beta)  + (1- \gamma_c) \mathcal{S}_r(\beta) $, which is the maximum $\SIR$ coverage in the interval $0 \leq \beta \leq \frac{R_L \tan \(\frac{\theta}{2}\)}{\L}$. \hfill $\blacksquare$

\begin{IEEEbiography}[{\includegraphics[width=1in,height=1.25in,clip,keepaspectratio]{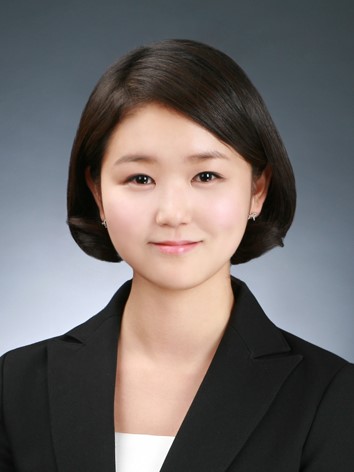}}]{Jeemin Kim} received the B.S. degree in electronic engineering from Ewha Womans University, Seoul, Korea, in 2012. She is currently working toward the combined Master's and Doctoral degrees in electrical and electronic engineering from Yonsei University, Seoul, Korea. 

She was a Visiting Doctoral Student with Wireless@KTH, Royal Institute of Technology, Kista, Sweden.
Her research interests include IoT communications, dynamic spectrum access, millimeter-wave communications, resource management, and stochastic geometric approach to analysis network interference.
\end{IEEEbiography}

\begin{IEEEbiography}[{\includegraphics[width=1in,height=1.25in,clip,keepaspectratio]{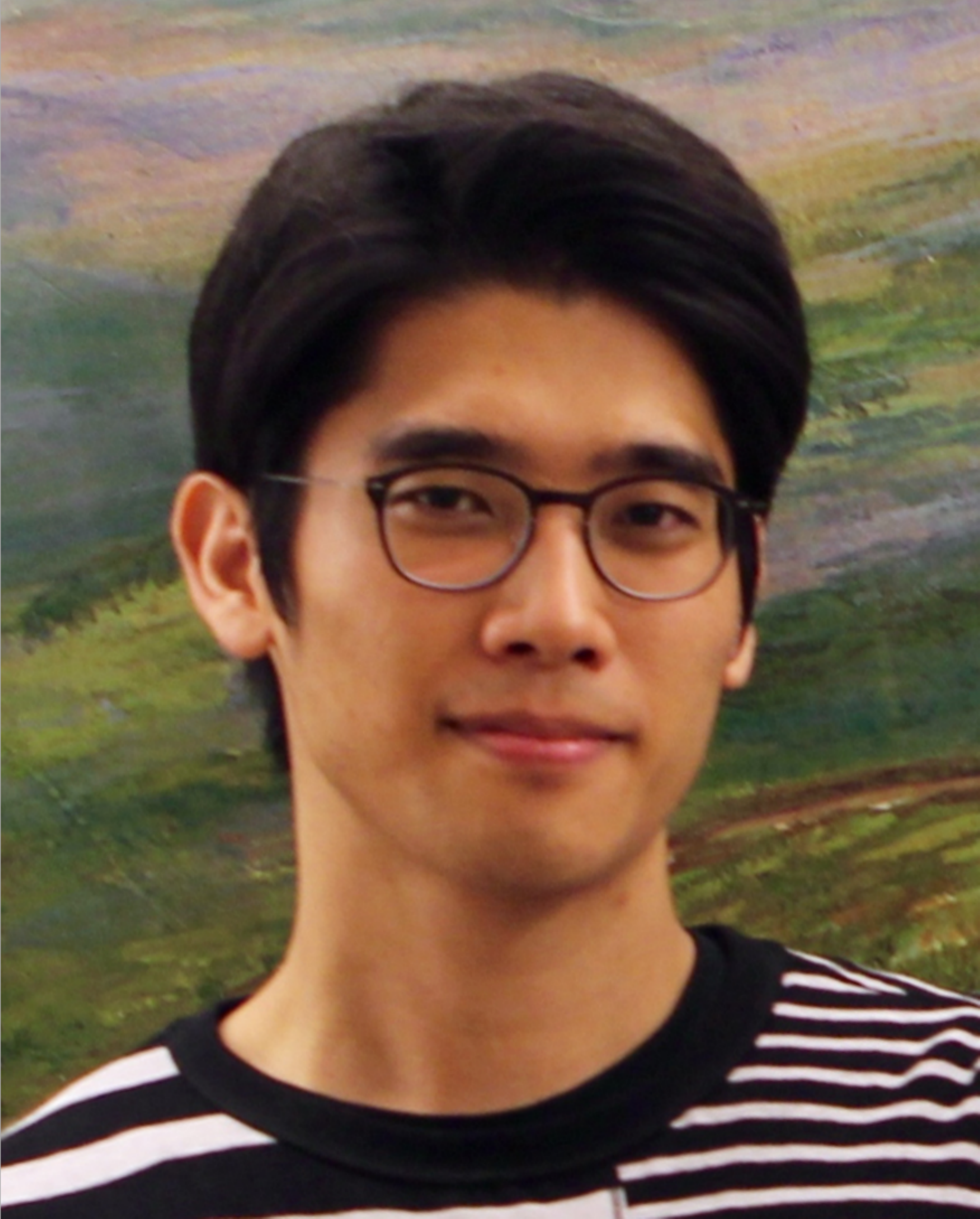}}]{Jihong Park} received his B.S. and Ph.D. degrees in electrical and electronic engineering from Yonsei University, Seoul, Korea, respectively in 2009 and 2016. He is currently a Postdoctoral researcher at Aalborg University, Denmark. He received the 2014 IEEE GLOBECOM travel grant, the 2014 IEEE Seoul Section Student Paper Contest Bronze Prize, and the 6th IDIS-ETNEWS (The Electronic Times) Paper
Contest Award sponsored by the Ministry of Science, ICT, and Future Planning of Korea. He was a visiting researcher at the Department of Applied Mathematics, Hong Kong Polytechnic University, at the Department of Communication Systems, the KTH Royal Institute of Technology, Stockholm, Sweden, at the Department of Electronic Systems, Aalborg University, Denmark, and at the Department of Electrical and Computer Engineering, New Jersey Institute of of Technology, USA, respectively in 2013, 2015, 2016, and 2017. His research interests include ultra-dense/ultra-reliable/massive-MIMO wireless system designs using stochastic geometry, network economics, and communication theory.
\end{IEEEbiography}

\begin{IEEEbiography}[{\includegraphics[width=1in,height=1.25in,clip,keepaspectratio]{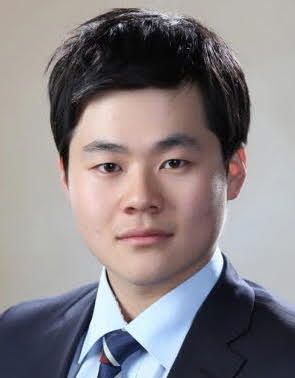}}]{Seunghwan Kim} received his B.S. degree in electrical and electronic engineering from Yonsei University, Seoul, Korea, in 2015. He is currently pursuing a combined Master's and Doctoral program in the School of Electrical and Electronic Engineering at the same university. His current research includes radio resource management, interference-limited networks, dual connectivity, power control, and network simulator implementation.
\end{IEEEbiography}

\begin{IEEEbiography}[{\includegraphics[width=1.05in,height=1.25in,clip,keepaspectratio]{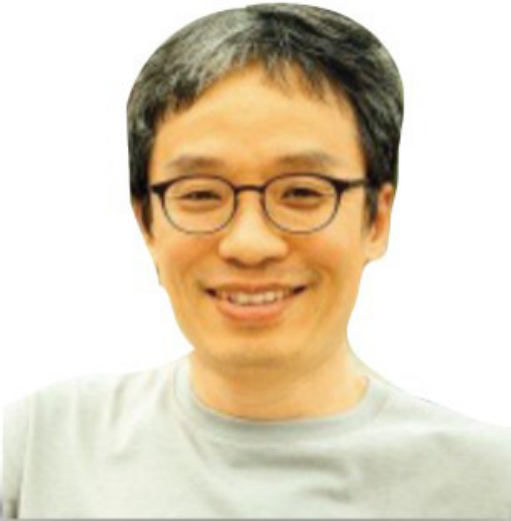}}]{Seong-Lyun Kim}
is a Professor of wireless networks at the School of Electrical \&
Electronic Engineering, Yonsei University, Seoul, Korea, heading the
Radio Resource Management \& Optimization  (RAMO) Laboratory and the
Center for Flexible Radio (CFR+). He was an Assistant Professor of
Radio Communication Systems at the Department of Signals, Sensors \&
Systems, Royal Institute of Technology (KTH), Stockholm, Sweden. He
was a Visiting Professor at the Control Group, Helsinki University
of Technology (now Aalto), Finland, and the KTH Center for Wireless
Systems. He served as a technical committee member or a chair for
various conferences, and an editorial board member of {\it IEEE
Transactions on Vehicular Technology}, {\it IEEE Communications
Letters}, {\it Elsevier Control Engineering Practice}, and {\it
Journal of Communications and Network}. He served as the
leading guest editor of {\it IEEE Wireless Communications}, and {\it
IEEE Network} for wireless communications in networked robotics. His
research interest includes radio resource management and information
theory in wireless networks, economics of wireless systems, and
robotic networks. He published numerous papers, including the
co-authored book (with Prof. Jens Zander), {\it Radio Resource
Management for Wireless Networks} (Artech House, Inc.).
\end{IEEEbiography}

\begin{IEEEbiography}[{\includegraphics[width=1in,height=1.25in,clip,keepaspectratio]{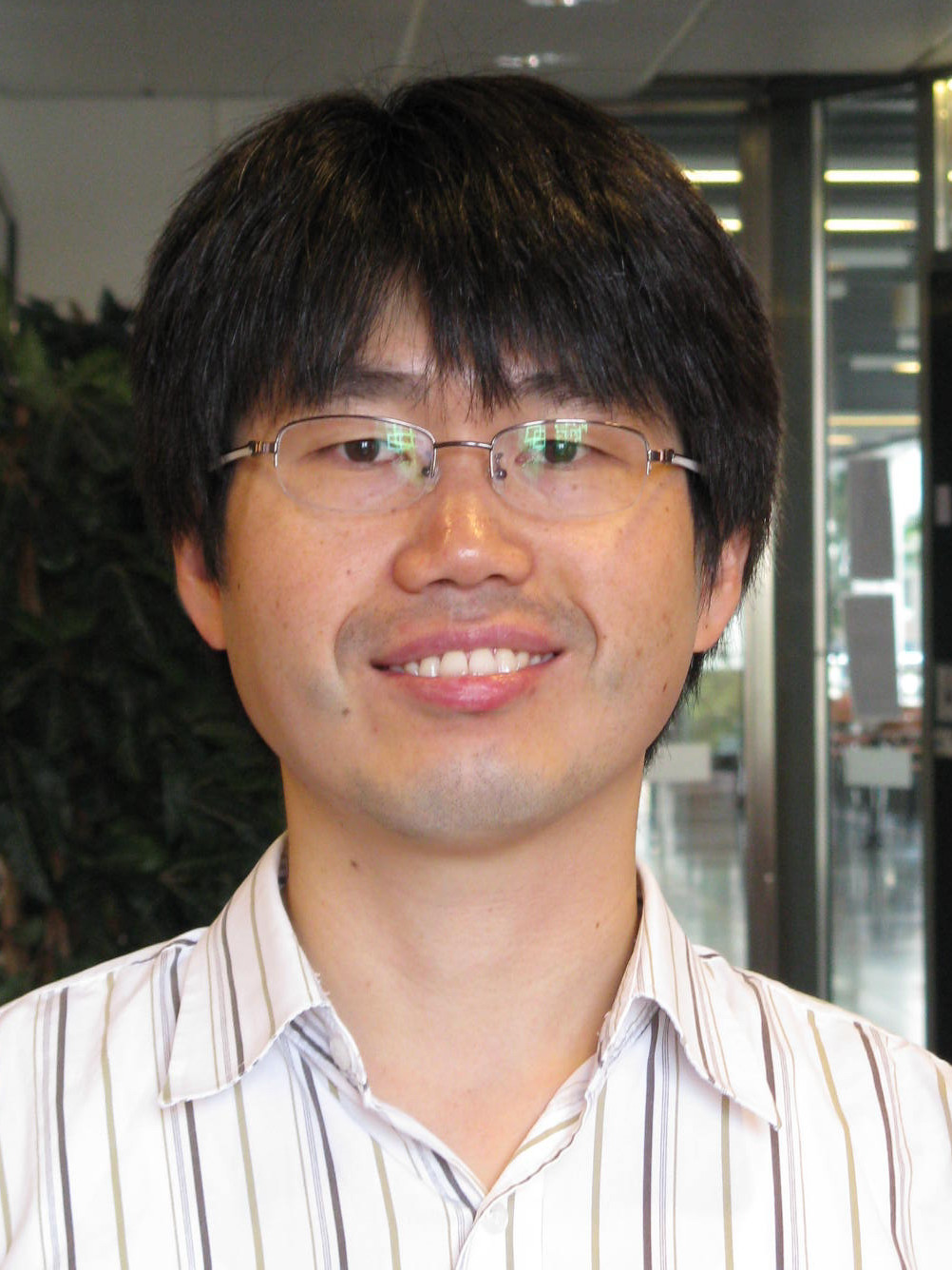}}]{Ki Won Sung}
(M'10) is a Docent researcher in the Communication Systems Department at KTH Royal Institute of Technology, Stockholm, Sweden. He is also affiliated with KTH Center for Wireless Systems (Wireless@kth). He received a B.S. degree in industrial management, and M.S. and Ph.D. degrees in industrial engineering from Korea Advanced Institute of Science and Technology (KAIST) in 1998, 2000, and 2005, respectively. From 2005 to 2007 he was a senior engineer in Samsung Electronics, Korea, where he participated in the development and commercialization of a mobile WiMAX system. In 2008 he was a visiting researcher at the Institute for Digital Communications, University of Edinburgh, United Kingdom. He joined KTH in 2009. He has participated in several European collaboration projects such as QUASAR, METIS, and METIS-II. His research interests include 5G technologies and architecture, energy-efficient wireless networks, and techno-economics of wireless systems.
\end{IEEEbiography}

\begin{IEEEbiography}[{\includegraphics[width=1in,height=1.25in,clip,keepaspectratio]{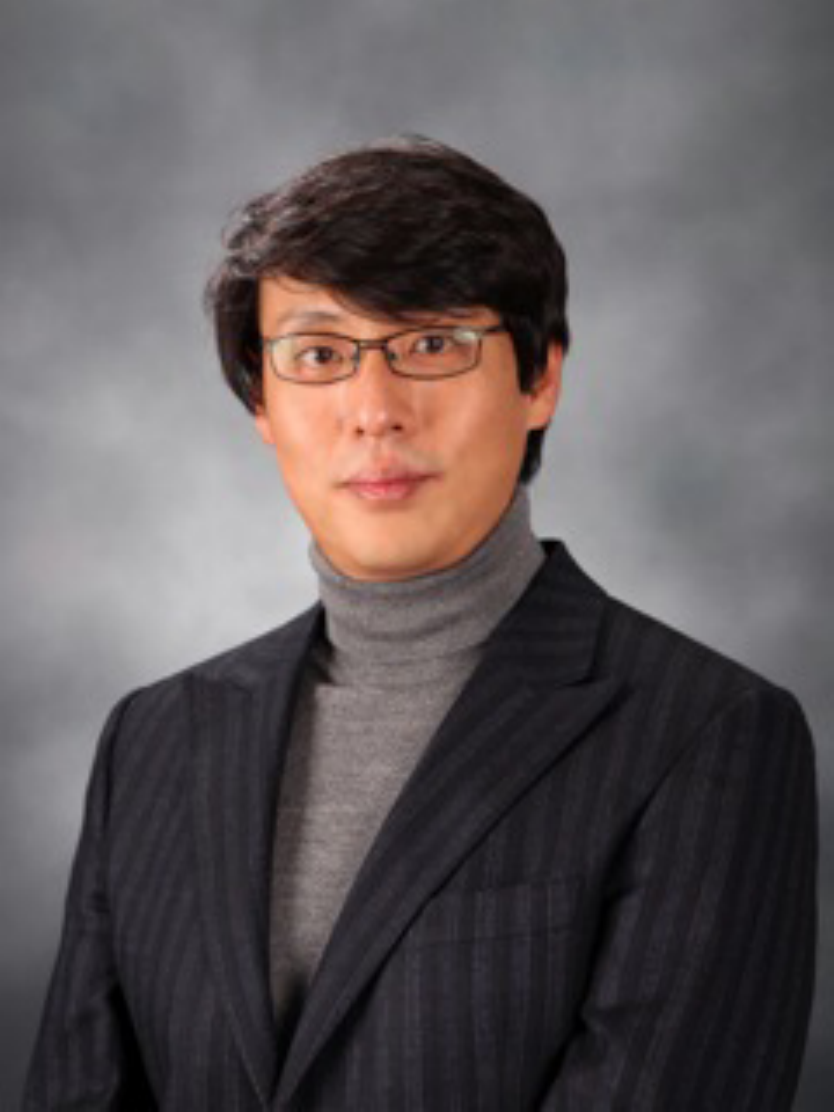}}]{Kwang Soon Kim}
(S'95, M'99, SM'04) was born in Seoul, Korea, on September 20, 1972. He
received the B.S. (summa cum laude), M.S.E., and Ph.D. degrees in
Electrical Engineering from Korea Advanced Institute of Science and
Technology (KAIST), Daejeon, Korea, in February 1994, February 1996,
and February 1999, respectively.\\
\indent From March 1999 to March 2000,
he was with the Department of Electrical and Computer Engineering,
University of California at San Diego, La Jolla, CA, U.S.A., as a
Postdoctoral Researcher. From April 2000 to February 2004, he was
with the Mobile Telecommunication Research Laboratory, Electronics
and Telecommunication Research Institute, Daejeon, Korea as a Senior
Member of Research Staff. Since March 2004, he has been with the
Department of Electrical and Electronic Engineering, Yonsei
University, Seoul, Korea, now is an Associate Professor.\\
\indent Prof. Kim is a Senior Member of the IEEE, served as an Editor of the Journal of the Korean Institute of Communications and Information Sciences (KICS) from 2006-2012, as the Editor-in-Chief of the journal of KICS since 2013, as an Editor of the Journal of Communications and Networks (JCN) since 2008, as an Editor of the IEEE Transactions on Wireless Communications since 2009.
\indent He was a recipient of the Postdoctoral Fellowship from Korea
Science and Engineering Foundation (KOSEF) in 1999. He received the
Outstanding Researcher Award from Electronics and Telecommunication
Research Institute (ETRI) in 2002, the Jack Neubauer Memorial Award
(Best system paper award, IEEE Transactions on Vehicular Technology)
from IEEE Vehicular Technology Society in 2008, and LG R\&D Award: Industry-Academic Cooperation Prize, LG Electronics, 2013.
His research interests are in signal processing, communication theory, information theory, and stochastic geometry applied to wireless heterogeneous cellular networks, wireless local area networks, wireless D2D networks and wireless ad doc networks.
\end{IEEEbiography}

\end{document}